\documentclass[reprint,
 amsmath,amssymb,
 aps,
 pra,
]{revtex4-2}
\usepackage{graphicx} 
\usepackage{amsmath}
\usepackage{amsthm}
\usepackage{mathtools}
\usepackage{physics}
\usepackage{dcolumn}
\usepackage{xcolor}
\usepackage{bm}
\usepackage{hyperref}
\usepackage[mathlines]{lineno}

\usepackage{tabularray}
\usepackage{diagbox}
\usepackage{array}

\usepackage{booktabs}

\newcommand{\rvec}[1]{\pmb{#1} }
\newcommand{\Cq}[0]{\mathbf{C^q} }
\newcommand{\Cc}[0]{\mathbf{C^c} }

\newsavebox{\mstrut}

\newcommand{\bbra}[1]{%
    \sbox{\mstrut}{\(#1\)}%
    \mathinner{\left\langle\kern-0.2\ht\mstrut\left\langle{#1}\right|\mkern-2mu\right|}%
}
\newcommand{\kett}[1]{%
    \sbox{\mstrut}{\(#1\)}%
    \mathinner{\left|\mkern-2mu\left|{#1}\right\rangle\kern-0.2\ht\mstrut\right\rangle}%
}
\newtheorem{theorem}{Theorem}
\newtheorem{definition}{Definition}

\newtheorem{proposition}{Proposition}
\newtheorem{lemma}{Lemma}

\begin{document}
\preprint{APS/123-QED}

\title{Dequantizing quantum machine learning models using tensor networks}

\author{Seongwook Shin}
\affiliation{Department of Physics and Astronomy, 
	Seoul National University, 08826 Seoul, South Korea}
 
\author{Yong Siah Teo}
\email{yong.siah.teo@gmail.com}
\affiliation{Department of Physics and Astronomy, 
	Seoul National University, 08826 Seoul, South Korea}
 
\author{Hyunseok Jeong}
\email{h.jeong37@gmail.com}
\affiliation{Department of Physics and Astronomy, 
	Seoul National University, 08826 Seoul, South Korea}

\date{\today}

\begin{abstract}
Ascertaining whether a classical model can efficiently replace a given quantum model---\emph{dequantization}---is crucial in assessing the true potential of quantum algorithms. In this work, we introduced the dequantizability of the function class of variational quantum-machine-learning~(VQML) models by employing the tensor network formalism, effectively identifying every VQML model as a subclass of matrix product state (MPS) model characterized by constrained coefficient MPS and tensor product-based feature maps. From this formalism, we identify the conditions for which a VQML model's function class is dequantizable or not. Furthermore, we introduce an efficient quantum kernel-induced classical kernel which is as expressive as given any quantum kernel, hinting at a possible way to dequantize quantum kernel methods. This presents a thorough analysis of VQML models and demonstrates the versatility of our tensor-network formalism to properly distinguish VQML models according to their genuine quantum characteristics, thereby unifying classical and quantum machine-learning models within a single framework.

\end{abstract}

\maketitle


\section{\label{Sec:I}Introduction}

Quantum machine learning (QML) garners a huge interest among various communities and industries in recent years as a prominent candidate for practical applications on quantum devices~\cite{Biamonte2017, Dunjko2020}. Variational QML (VQML) uses a variational quantum circuit as a data processor, and the variational parameters in the quantum circuit are optimized with the help of classical optimization algorithms in order to learn and predict data outputs. VQML aims to achieve a more powerful ML model by exploiting possible quantum advantage of quantum circuits in noisy intermediate scale quantum (NISQ) era.

While there exist theoretical proofs that demonstrate the possibility of achieving a quantum advantage in ML tasks in fully quantum settings \cite{Huang2022:QA, Huang2021:PRL}, more effort is required to understanding whether ML from classical data can also achieve such a quantum advantage~\cite{Ciliberto2018:QMLclassical, Alvi2023:JHEP, Monaco2023:QPD, Wei2023, D2CS00203E}.

For this purpose, a fair assessment of VQML and classical ML models is in order, both of which possess inherently different structures. Moreover, the pre-processing of classical data always precedes VQML when they are encoded on NISQ machines. This additional computation might lead to the ``dequantization" argument when comparing a classical and quantum model~\cite{Tang2021,Tang2022:dequantreview}. Moreover, if one does not have access to a coherent quantum memory and quantum channel, then even if the QML uses a `quantum state' as its input, one cannot avoid using classical data to `upload' the quantum state onto the quantum circuit. In this study, we propose a unified tensor-network (TN) formalism to systematically analyze VQML models, which permits us to classify all classical-data-encoded VQML models into a subclass of matrix product state (MPS) ML models~\cite{Stoudenmire2016supervised}. We introduce the concept of dequantization of the function class of VQML models---the efficient approximation of all function-class outputs of a VQML model using a classical model---and find necessary conditions for (non-)dequantizable VQML models by classical MPS models.

More specifically, the TN formalism describes the function output of a VQML model as a \textit{linear} MPS model form, subsequently separating it into two components: the coefficient part of the linear model which is in the form of MPS containing all quantum-circuit training parameters and the basis part (or a feature map in the ML lingo), which formulates the basis for the linear model. The number of linearly independent basis functions can scale exponentially with the number of encoding gates~\cite{Shin2023, Evan2022:benign}, challenging classical models to approximate them. However, by leveraging the knowledge of data pre-processing before implementing VQML, we can simply observe that the basis part is in an easily manageable tensor-product form. 

Representing coefficients as an MPS allows systematic analysis of expressivity and approximability of models in the context of entanglement. Moreover, we discover that the coefficients of a VQML model are Pauli coefficients of the circuit-dependent operator when expanded in the Pauli basis. This observation enables systematic analysis of coefficients of VQML models using various techniques, of which we shall provide some hints.  

To assess whether a VQML model is dequantizable or not, we construct a classical MPS model having the same basis function as (or is \emph{basis-equivalent} to) the VQML model and explore the possibility of VQML function-class dequantization. With dimensional arguments and borrowing key results concerning MPS approximability~\cite{Verstraete2006:MPSseminal}, we list some necessary conditions of non-dequantizable VQML models. These include models with dimensions that scale exponentially with the number of qubits, and coefficient MPSs that are highly entangled. Numerically we show that a general poly-depth variational quantum circuit with a non-trivial encoding strategy can satisfy this requirement.

Lastly, we introduce the computationally efficient classical kernel inspired by the basis-equivalent classical MPS model that is naturally attained by using the equivalent pre-computations as the given quantum kernel. It covers the function space from the quantum kernel, and we compare the performance of the classically hard-to-simulate quantum kernel and the classical counterpart of it. 

After a preliminary outline of the theoretical background and setup of VQML models in Sec.~\ref{Sec:II}, the unifying tensor-network formalism for describing these quantum models is introduced in Sec.~\ref{Sec:III}, followed by a more detailed discussion of VQML dequantization in Sec.~\ref{Sec:IV}. In Sec.~\ref{Sec:V}, upon recognizing that the feature map is, in fact, efficient to handle classically, we analytically and numerically study basis-equivalent classical MPS models and identify conditions for (non-)dequantizable VQML models. Then, using our TN formalism, in Sec.~\ref{Sec:VI}, we construct tensor-product classical kernel models and show that they can efficiently cover the quantum kernel counterparts. This work shall finally conclude in Sec.~\ref{Sec:VII}.


\section{\label{Sec:II}Preliminaries of variational quantum machine learning model} 

\begin{figure*}[ht]
    \centering
     \includegraphics[width = 0.95\textwidth]{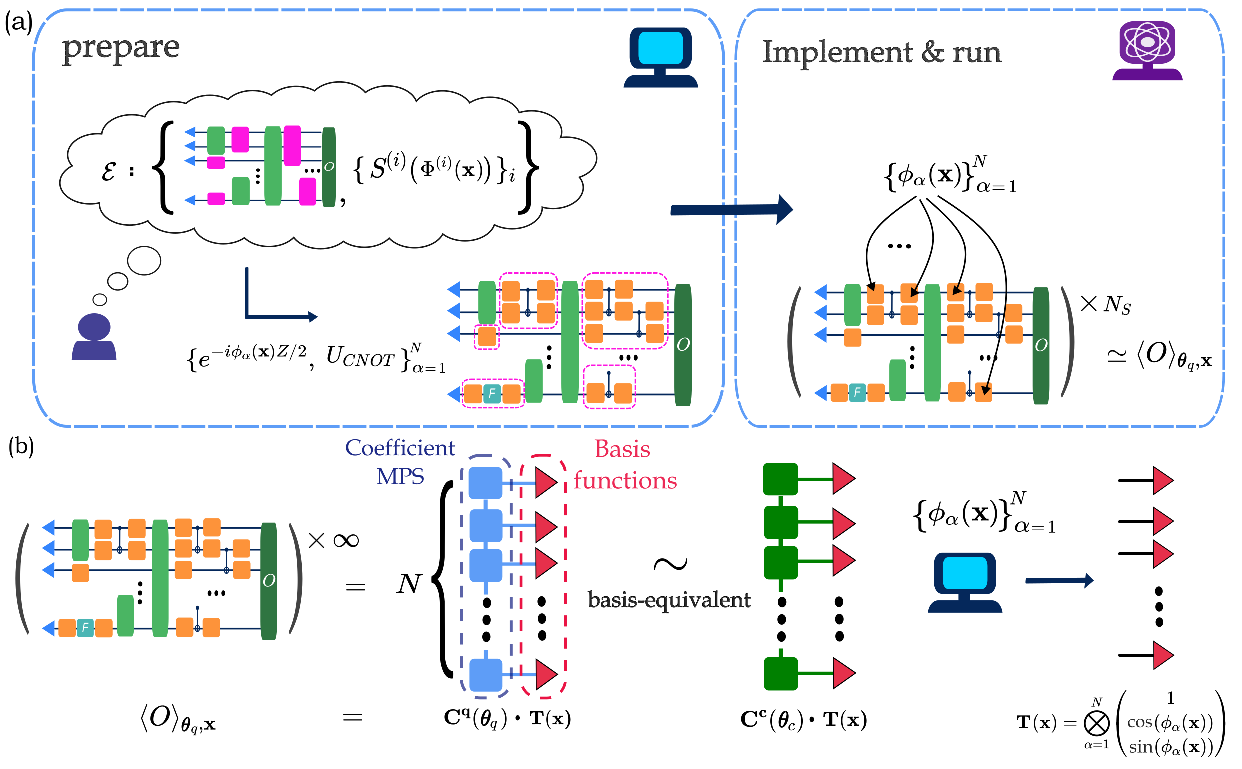}
     \caption{Schematic overview of this work. (a) The procedure for Variational Quantum Machine Learning (VQML) with a general encoding strategy, denoted as $\mathcal{E}$. This strategy includes encoding gates $S^{(i)}(\cdot)$s, pre-processing functions $\Phi^{(i)}(\rvec{x})$, and their respective positions within the quantum circuit (represented by magenta boxes). During the preparation stage, all original encoding gates are compiled into single-qubit Pauli-Z rotations (represented by orange boxes), with angles $\phi_{\alpha}(\rvec{x})$s and non-parametrized two-qubit gates, all computed classically. This decomposition may include non-parametrized unitaries, denoted as $F$. Green boxes represent the trainable circuit with variable parameters $\rvec{\theta}_q$. The VQML-model output is given by the expectation value of a specific observable $O$. This output is estimated \textit{via} $N_S$ runs of the quantum circuit. (b) The exact value of the VQML model can be represented as a linear model using the feature map $\mathbf{T}$ and constrained MPS coefficient tensor $\Cq(\rvec{\theta}_q)$, constructed from the parametrized circuit. With the pre-processing functions obtained during the preparation stage, we can efficiently construct $\mathbf{T}$, classically. The classical Tensor Network (TN) model using $\mathbf{T}$ resides in the same function space spanned by the same basis functions as the VQML model. The TN formalism can then be used to compare the respective MPS parts $\Cc(\rvec{\theta}_q)$ and $\Cq(\rvec{\theta}_q)$ for the classical and quantum models, which dictates the possibility for dequantization. \label{fig:fig1}}
\end{figure*}

Machine Learning (ML) can be understood as a function approximation task, where the target function is unknown and is to be learned from a training dataset. A function approximator in ML is a computational model, which defines and generates some function class. The cost function measures how well our function-approximator model is approximating the target function. An ML algorithm minimizes this cost function calculated with the training dataset and ML model function, by using various numerical or analytical methods.

Variational quantum machine learning (VQML) is ML that uses a parametrized quantum circuit as a computational model. The parametrized quantum circuit usually has a fixed structure (called an \emph{ansatz}) and is parametrized by variable parameters. Because VQML models employ quantum circuits, they take quantum state as input, which inevitably requires a classical-to-quantum encoding procedure~\footnote{This classical-to-quantum encoding procedure is even necessary for some QML tasks that use the quantum state as its input. This is the case where quantum states are stored in the classical form and re-constructed on the quantum computer later using classical information. In fact, if there is no coherent quantum memory and channel, no QML can avoid the classical-to-quantum encoding process.}. There exist various encoding strategies, such as amplitude encoding, Pauli encoding, data re-uploading~\cite{Perez2020}, instantaneous quantum polynomial (IQP) encoding that is conjectured to be hard to simulate classically~\cite{Havlicek2019}, and so on. These encoding strategies, $\mathcal{E}$, consists of pre-processing functions $\Phi^{(i)} : \mathbb{R}^d \rightarrow \mathbb{R}^{m_i^2-1}$, $m_i$-qubit encoding gates $S^{(i)}(\cdot) : \mathbb{R}^{m_i^2-1} \rightarrow \mathbb{C}^{2^m_i}$ that map pre-processed data $\Phi^{(i)}(\rvec{x})$ into an $m_i$-qubit state, and positions of encoding gates within the quantum circuit. Here we assumed that the inputs are $d$-dimensional real vectors without loss of generality and $i$ is the index for distinguished encoding gates. For a general $\mathcal{E}$, the corresponding encoding gates $S^{(i)}$s can be highly non-local.

To the output of the VQML model, we choose some observable (or POVM) $O$ and measure its expectation value. Then, the function class of the VQML model is defined as
 \begin{equation}\label{eq:generalVQML}
     f_Q(\rvec{x};\mathcal{E}, U, \rvec{\theta}, O) = \bra{\mathbf{0}}U^{\dag}(\rvec{x};\mathcal{E},\rvec{\theta})OU(\rvec{x};\mathcal{E},\rvec{\theta})\ket{\mathbf{0}}.
 \end{equation} Here, we initiate the $n$ qubits to the state $\ket{\mathbf{0}} \equiv \ket{0}^{\otimes n}$, and $U(\rvec{x}; \mathcal{E}, \rvec{\theta}$) represents the quantum circuit using encoding strategy $\mathcal{E}$  and trainable unitaries which are parametrized by $\rvec{\theta}$.

Let us consider the general $\mathcal{E}$, where $S^{(i)}$ are multi-qubit gates. To implement any $m_i$-qubit encoding gate on a real quantum circuit, we need to compile it using a universal gate set of the quantum device. Here we assume a universal gate set comprising an arbitrary single-qubit gate and some unparametrized two-qubit gate such as a controlled-not~(CNOT) gate. As arbitrary universal gate sets can be converted to this single- and CNOT gate set, without loss of generality, every multi-qubit encoding gate is decomposed with a set $\{ S_1^{\alpha}(\tilde{\phi}^{\alpha}(\rvec{x})), U_{\text{CNOT}}\}_{\alpha}$ before running VQML algorithm. Single-qubit gates $S_1^{\alpha}$s are again decomposed to Pauli rotation gates,
 \begin{align}
    S_1^{\alpha}(\tilde{\phi}^\alpha(\rvec{x})) &= e^{-i \dfrac{\phi^{\alpha}_1(\rvec{x})}{2}Z} e^{-i \dfrac{\phi^{\alpha}_2(\rvec{x})}{2}Y} e^{-i \dfrac{\phi^\alpha_3(\rvec{x})}{2}Z}\\
    & = e^{-i \dfrac{\phi^\alpha_1(\rvec{x})}{2}Z}F^{\dag} e^{-i \dfrac{\phi^\alpha_2(\rvec{x})}{2}Z} F e^{-i \dfrac{\phi^\alpha_3(\rvec{x})}{2}Z},
 \end{align}
where $F = \dfrac{1}{\sqrt{2}}\begin{pmatrix}
    &1 &1\\ &i &-i
\end{pmatrix}$, such that $F^{\dag} Y F = Z$, and $\phi^{\alpha}_1, \phi^{\alpha}_2,~\text{and} \phi^{\alpha}_3 $ are pre-processing functions that are obtained when compiling multi-qubit encoding gates into the Pauli-rotation form. We denote $N$ as the total number of single-qubit Pauli-Z rotation gates when all the encoding gates are compiled. For simplicity, we combine upper and lower indices in $\phi_k^\alpha$ into one index $\alpha \in [N]$, and group all pre-processing functions $\phi_{\alpha}(\mathcal{E}) : \mathbb{R}^d \rightarrow \mathbb{R}$ and the values $\{\phi_\alpha(\rvec{x}; \mathcal{E})\}_{\alpha=1}^N$\mbox{[see Fig.~ \ref{fig:fig1}(a)]}. 

\section{\label{Sec:III}The Function class of VQML models}

After the Pauli-gate decomposition, all data-dependent encoding gates are expressed in terms of Pauli-Z rotations. Using the result from \cite{Schuld2021:Fourier}, it is straightforward to see that the function class of a general encoding strategy corresponds to a linear combination of basis functions $\{B_j(\rvec{x}; \mathcal{E})\}_j$,
\begin{equation}
 \begin{split}
     f_Q(\rvec{x};\rvec{\theta}, \mathcal{E}, U, O) &=\sum_{j =1}^{K} c_j(\rvec{\theta}, U, O) e^{-ib_j(\rvec{x};\mathcal{E})} \\
     &\equiv \sum_{j=1}^{K} c_j(\rvec{\theta}, U, O) B_{j}(\rvec{x};\mathcal{E}) \\
     & \equiv \rvec{c}(\rvec{\theta}, U, O) \cdot \mathbf{B}(\rvec{x}; \mathcal{E}),
 \end{split}\label{eq:FLM}
 \end{equation}
where 
 \begin{equation}\label{eq:basis}
     b_j(\rvec{x}; \mathcal{E}) \in \left\{\sum_{\alpha=1}^{N} \beta_\alpha \phi_{\alpha}(\rvec{x}; \mathcal{E}) \,\vert\, \beta_\alpha = \{-1, 0, 1\} \right\}.
\end{equation} The symbol $K \leq 3^N$ refers to the number of linearly independent basis functions. In other words, any VQML model is a featured linear model (FLM) which is a linear model in the feature space endowed by a feature map $\mathbf{B} : \mathbb{R}^d \rightarrow \mathbb{C}^K$~\cite{Havlicek2019, Schuld2019:QMLinFeature}. Note that the coefficients $c_j(\rvec{\theta}, U, O)$s from the quantum model are not arbitrary, but constrained as they are obtained from a quantum circuit. Calculating the exact form or values of $c_j(\rvec{\theta}, U, O)$s is equivalent to simulating a quantum circuit directly, which is typically inefficient unless the circuits possess special structures~\cite{Pashayan2020fromestimationof}. Rather, we shall analyze VQML models using a unifying TN framework, to be introduced in the following subsection.

\subsection{\label{sec:IIIA}VQML models as matrix product state models}

MPS model is a variational ML model which is a featured linear model. A feature map is given by a tensor-product of certain data-dependent vectors and a coefficient part is given by a variational MPS. The MPS model was originally introduced in Ref.~\cite{Stoudenmire2016supervised} as a quantum-inspired classical model. However, here we assert a somewhat ``reverse" statement that a VQML model using classical data is a subclass of the MPS model.

Throughout this text, we assume all encoding gates are transformed to single-qubit Pauli-Z rotations. Additionally, the set of pre-processing functions $\{\phi_\alpha\}_{\alpha=1}^N$ are defined as in Sec~\ref{Sec:II}. We omit the encoding strategy $\mathcal{E}$ dependency for notational simplicity. 

First, let us consider the simple parallel VQML model with $n = N$ where all the encoding gates are placed parallel and in between trainable unitaries $W_1(\rvec{\theta}_1)$ and $W_2(\rvec{\theta}_2)$:
\begin{equation}
    \begin{split}
    f_Q(\rvec{x};\rvec{\theta},W_1,W_2, O) &= \bra{0}W_1^{\dag}(\rvec{\theta}_1)\mathbf{S}^{\dag}(\rvec{x}) W_2^{\dag}(\rvec{\theta}_2) O \\& W_2(\rvec{\theta}_2)\mathbf{S}(\rvec{x}) W_1(\rvec{\theta}_1)\ket{0}, \label{eq:simple parallel}
\end{split}
\end{equation}
where $\mathbf{S}(\rvec{x}) = \Pi_{\alpha=1}^N e^{-i\phi_\alpha(\rvec{x})Z_\alpha/2}$, and $\rvec{\theta} \equiv (\rvec{\theta}_1, \rvec{\theta}_2)$. Then the following lemma holds, where detailed proof with graphical description is given in Appendix.~\ref{app:A}.

\begin{lemma}[A simple parallel VQML model is a MPS model\label{lem:simpleparallel}]
    Given a simple parallel VQML model as Eq.~\eqref{eq:simple parallel}, one can represent it as:
    \begin{equation}
        f_Q(\rvec{x};\rvec{\theta},W_1,W_2,O) = \Cq(\rvec{\theta})\cdot \mathbf{T}(\rvec{x}),
    \end{equation}
    with the \textit{coefficient MPS}
    \begin{equation}
        \mathbf{C^q}(\rvec{\theta}) = (O'\odot \rho^T)(\rvec{\theta}) \cdot \mathbf{\tilde{P}},
    \end{equation}
    and
    \begin{equation}
    \mathbf{\tilde{P}} = \begin{pmatrix}
        1&0&0\\
        0&1&i\\
        0&1&-i\\
        1&0&0
    \end{pmatrix}^{\otimes N}.
\end{equation}
    The \textit{feature map is given as}
    \begin{equation}
    \mathbf{T}(\rvec{x}) = \bigotimes_{\alpha=1}^N\mathbf{T}^{(\alpha)}(\rvec{x}) = \bigotimes_{\alpha=1}^N\begin{pmatrix}
        1\\
        \cos{(\phi_\alpha(\rvec{x}))}  \\
        \sin{(\phi_\alpha(\rvec{x}))}  \\
    \end{pmatrix}.
\end{equation}
\end{lemma}
 We denoted the evolved observable as $O'(\rvec{\theta}_2) := W_2^{\dag}(\rvec{\theta}_2) O W_2(\rvec{\theta}_2)$, pre-encoded state as $\rho(\rvec{\theta}_1) := W_1(\rvec{\theta}_1)\ketbra{0}W_1^{\dag}(\rvec{\theta}_1) $, tensor contraction as $\cdot$, and the Hadamard product as $\odot$. The tensor $(O'\odot \rho^T)(\rvec{\theta})$ is a $2^N \times 2^N$ matrix having $2N$ indices, where row and column indices are decomposed into $N$ indices each for the one-qubit line. See Fig.~\ref{fig:TNbending}(a). We \textit{vectorize} this $(O' \odot \rho^{T})(\rvec{\theta})$ by gathering the same site indices to make it a tensor of $N$ indices having a dimension of $4$. This enables contraction between tensor network $\mathbf{\tilde{P}}$ and $(O'\odot\rho^T)$.

Next, we consider the general case of a VQML model using $n$ qubits as given in Eq.~\eqref{eq:generalVQML}. We can rewrite any general encoded VQML as
\begin{equation}\label{eq:generalunravel}
    \begin{split}
     f_Q(\rvec{x};\rvec{\theta})& = \bra{\rvec{0}}W_0^{\dag}(\rvec{\theta}_0)\mathcal{S}_1^{\dag}(\rvec{x}) W_1^{\dag}(\rvec{\theta}_1)\mathcal{S}_2^{\dag}(\rvec{x})\cdots \\ &S^{\dag}_R(\rvec{x})W^{\dag}_R(\rvec{\theta}_R) O W_R(\rvec{\theta}_R)\mathcal{S}_R(\rvec{x}) \cdots \\ &\mathcal{S}_2(\rvec{x})W_1(\theta_1)\mathcal{S}_1(\rvec{x}) W_0(\rvec{\theta}_0)\ket{\rvec{0}},
    \end{split}
\end{equation}
This equation distinguishes layers by their dependencies on data or variational parameters, aligning with the concept of a data re-uploading model as noted in Ref.~\cite{Perez2020}. Recall that all encoding gates are decomposed to single-qubit Pauli Z gates and the pre-processing functions $\{\phi_\alpha\}_{\alpha}$ are given. Equation~\eqref{eq:generalunravel} includes cases where some $\mathcal{S}_k$ contains the trivial pre-processing function $\phi(\bm{x})=0$. Leveraging lemma~\ref{lem:simpleparallel}, we can now establish a theorem regarding general VQML models.

\begin{theorem}[\label{thm:1}Any VQML model is a MPS model]
    Any VQML model employing a general encoding strategy $\mathcal{E}$, and a circuit ansatz $U$ such as Eq.~\eqref{eq:generalVQML} can be represented as an MPS model:
    \begin{equation}
        f_Q(\rvec{x};\mathcal{E},U,\rvec{\theta},O) = \Cq(\rvec{\theta})\cdot\mathbf{T}(\rvec{x}),
    \end{equation}
    where
    \begin{equation}
        \mathbf{C^q}(\rvec{\theta}) = (O'_R\odot \rho_R^T)(\rvec{\theta}) \cdot \mathbf{\tilde{P}},
    \end{equation}
    and
    \begin{equation}
    \mathbf{T}(\rvec{x}) = \bigotimes_{\alpha=1}^{nR}\mathbf{T}^{(\alpha)}(\rvec{x}) = \bigotimes_{\alpha=1}^{nR}\begin{pmatrix}
        1\\
        \cos{(\phi_\alpha(\rvec{x}))}  \\
        \sin{(\phi_\alpha(\rvec{x}))}  \\
    \end{pmatrix}.
    \end{equation}
    In this formulation, the circuit ansatz $U$, $O$ and $\rvec{\theta}$ dependent tensors $O'_R$ and $\rho_R$ are as specified in Eq.~\eqref{eq:app:generalO} and Eq.~\eqref{eq:app:generalrho} respectively.    
\end{theorem}
The proof of Theorem~\ref{thm:1} with graphical description can be found in Appendix.~\ref{app:A}

The function class of a VQML model is a linear model in the feature space, with the feature map $\mathbf{T}$. This can also be viewed as a special MPS model, characterized by a special coefficient MPS $\Cq(\rvec{\theta})$, which is determined by the quantum circuit's structure. In the following sections, we delve deeper into the analysis of both $\mathbf{T}$ and $\Cq(\rvec{\theta})$. This exploration aims to reveal the insights and implications that such an analysis can provide. 

\subsection{\label{Sec:IIIB}The feature map $\mathbf{T}$}

The feature map $\mathbf{T} : \mathbb{R}^d \rightarrow \mathbb{R}^{3^N}$ (where $N$ is now the length of the coefficient MPS that depends on the structure of the VQML model) is a mapping given by 
\begin{equation}
    \mathbf{T}(\rvec{x}) = \bigotimes_{\alpha=1}^N\mathbf{T}^{(\alpha)}(\rvec{x}) = \bigotimes_{\alpha=1}^N\begin{pmatrix}
        1\\
        \cos{(\phi_\alpha(\rvec{x}))}  \\
        \sin{(\phi_\alpha(\rvec{x}))}  \\
    \end{pmatrix}.
\end{equation}
The encoding strategy $\mathcal{E}$ determines pre-processing functions $\{\phi_\alpha\}_\alpha$, consequently defining the feature map. Output functions of the VQML model are given by the linear sum of basis functions in $\{\mathbf{T}_i(\rvec{x})\}_i^{3^N}$, which is the set of components of the feature map $\mathbf{T}(\rvec{x})$. This means the function space, or the function class, of the VQML model, is spanned by these $3^N$ functions. It is critical to recognize that these basis functions may not all be linearly independent, as their independence hinges on the selected pre-processing functions and, therefore, $\mathcal{E}$. 

As an example of a 1D VQML function, in a naive Pauli encoding strategy where all $\phi_\alpha(x) = x$, only $2N+1$ out of $3^N$ components are linearly independent. On the other hand, for the exponential encoding strategy~\cite{Shin2023} where $\phi_\alpha(x) = k^{\alpha-1}x$, a set of $3^N$ linearly independent basis functions can be generated with $k\geq3$.

In terms of computational complexity, $\mathbf{T}(\rvec{x})$ is efficient to store and generate classically, as it requires only $O(N)$ memory to store and single call of $\phi_\alpha$s as the VQML model does. With this observation, we conclude that exponentially large feature space, commonly dubbed as a special feature of QMLs, is not unique to quantum models. 

\subsection{\label{Sec:IIIC}The coefficient MPS $\Cq$ }

The coefficient MPS $\Cq: \mathbf{\Theta} \times \mathbb{O} \rightarrow \mathbb{R}^{3^N}$ (For the general encoding case, $O'_R$ and $\rho_R$ respectively.) is a mapping from parameter space $\mathbf{\Theta}$ and observable space $\mathbb{O}$ to $3^N$-dimensional real space. This \textit{coefficient MPS} of given VQML, calculated as
\begin{equation}\label{eq:Cq}
\mathbf{C^q}(\rvec{\theta}) = (O'\odot \rho^T)(\rvec{\theta}) \cdot \mathbf{\tilde{P}},
\end{equation}
can be seen as a normal vector of hyperplane on the feature space.

Unlike coefficients in a simple linear model, we cannot control all $3^N$ components in $\Cq$ freely, as they are obtained implicitly by the contraction of unitaries in the quantum circuit. In general, it is not \textit{universal}, which means that not all of $3^N$ dimensional vectors can be generated. For $\Cq$ to be universal (besides the normalization condition), one needs universal \textit{ansatz} in trainable unitary parts and multiple circuits as one unitary orbit of the Hermitian matrix cannot cover the whole space of Hermitian matrix space.

In Eq.~\eqref{eq:Cq}, one might wonder what $\mathbf{\tilde{P}}$ does. This tensor is a  given by
\begin{equation}
    \begin{split}
        \mathbf{\tilde{P}} &= \begin{pmatrix}
            1&0&0\\
            0&1&i\\
            0&1&-i\\
            1&0&0
        \end{pmatrix}^{\otimes N}\\
        &= \left ( \kett{I}\bra{0} + \kett{X}\bra{1} + \kett{Y}\bra{2}\right)^{\otimes N}.
    \end{split}
\end{equation}
In other words, the coefficient tensor $\Cq$ is obtained by projecting out $Z-$containing Pauli coefficients of $O'\odot\rho^T$. Here, $\lambda_{\rvec{i}}$'s are the Pauli coefficients when $O'\odot\rho^T$ is represented with the Pauli basis as
\begin{equation}
    (O'\odot\rho^T)(\rvec{\theta}) = \sum_{\rvec{i}} \lambda_{\rvec{i}}(\rvec{\theta}) ~\sigma^{(1)}_{i_1}\otimes \sigma^{(2)}_{i_2}\otimes \cdots\otimes \sigma^{(N)}_{i_N}, 
\end{equation}
where $\rvec{i} = \{0,1,2,3\}^{\otimes N}$ and $\sigma_{i_k} = \{I,X,Y,Z\}$.
One can identify 
\begin{equation}\label{eq:coeff}
    \begin{split}
\mathbf{C^q}(\rvec{\theta})_{\Tilde{\rvec{i}}} &= ((O'\odot \rho^T)(\rvec{\theta}) \cdot \mathbf{\tilde{P}})_{\Tilde{\rvec{i}}}\\
&= 2^N\lambda_{\Tilde{\rvec{i}}},
    \end{split}
\end{equation}
where now we have truncated indices $\Tilde{\rvec{i}} \in \{0,1,2\}^{\otimes N}$.

For instance, consider the exponential encoding on 1d input $x$, where $\phi_\alpha(x) = 3^{\alpha-1}x$ and $N=3$. Then, one of the basis functions is $\cos(x)\sin(3x)\cos(9x)$, corresponding to $\Tilde{\rvec{i}} = (1,2,1)$. Therefore the coefficient of $\mathbf{T}_{121}(x) = \cos(x)\sin(3x)\cos(9x)$ is $2^N$ times that of the Pauli string $X\otimes Y \otimes X$ when $O'\odot \rho^T$ is represented in the Pauli basis.

This observation clarifies the previously opaque nature of a VQML model's coefficients and gives us a new way to understand the model in the context of operator spreading~\cite{Khemani2018:spreading, Xu2020:scrambling} or non-Cliffordness of the circuit~\cite{Leone2022:SRE}. We refer the Reader to Appendix~\ref{app:coefficients} for more detailed discussions. It also allows for a more comprehensive analysis of noise effects on VQML models using techniques introduced in Refs.~\cite{Aharonov2022:polysim,Fontana2023:ClassicSim}. Refer to Appendix~\ref{app:noisy} for additional information.

 \begin{figure}
 \centering
 \includegraphics[width = 0.95\columnwidth]{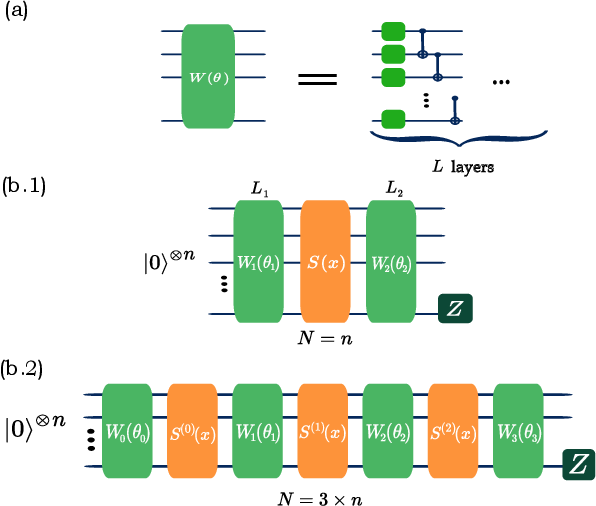}
    \caption{The schematic diagrams illustrate the VQML models used for simulations. (a) The structure of trainable unitaries. These consist of $L$ layers of hardware-efficient \textit{ansatze}. Each small green box represents a parametrized single-qubit unitary $U(\theta^{(1)},\theta^{(2)},\theta^{(3)})= \begin{pmatrix}
\cos(\theta^{(1)}/2) & -e^{i\theta^{(3)}}\sin(\theta^{(1)}/2)\\
e^{i\theta^{(2)}}\sin(\theta^{(1)}/2) & e^{i(\theta^{(2)} + \theta^{(3)})}\cos(\theta^{(1)}/2)
\end{pmatrix}$, which contains three free parameters. Though numerous options exist, This work focuses on results from this specific trainable circuit \textit{ansatz}, chosen as an illustrative example for comparing VQML and classical models within the TN formalism. (b.1) The diagrams for the simple parallel model. In this model, the number of qubits in the circuit, $n$, equals the number of single-qubit Pauli-Z encoding gates $N$. (b.2) The general structure (data re-upload) model. For the numerical results in this work, we consider a model with 3 times re-uploading. Hence, $N = 3n$.\label{fig:vqml models} }
\end{figure}

To fully contract the function $\Cq(\rvec{\theta})\cdot\mathbf{T}(\rvec{x})$ to get the output from the quantum model, computational resources on the order of $O(3\chi_q(\rvec{\theta})^2N)$ are required. Here, $\chi_q(\rvec{\theta},O) := \underset{k\in[N-1]}{\max} \chi^{(k)}_q$ represents the maximum bond dimension among all bond indices of $\Cq(\rvec{\theta})$. This highlights the significance of the maximum bond dimension in MPS, as it directly influences the computational complexity of contracting MPS. 

The value of $\chi_q$ depends on the circuit \textit{ansatz}. However, for VQML models that include multiple two-qubit entangling layers, which is a common \textit{ansatz} for variational circuits, $\chi_q$ can increase exponentially with the number of layers. To be more concrete, let us consider the `simple parallel model', depicted in Fig.~\ref{fig:vqml models}(a) and 2(b.1). Our trainable \textit{ansatz} $W_1(\rvec{\theta}_1)$ ($W_2(\rvec{\theta}_2)$) is `hardware-efficient' \textit{ansatz} which is composed of $L_1$ ($L_2$) layers of $N$ parametrized single-qubit unitaries and nearest-neighbor CNOT gates. That is
\begin{equation}
\begin{split}
    W_k(\rvec{\theta}) = \Pi_{l=1}^{L_k}(\Pi_{i=1}^{N-1} &U_{\text{CNOT}}^{l,i,i+1}\\&\times\Pi_{i=1}^{N} U^{(l,i)}(\theta^{(l,i)}_1, \theta^{(l,i)}_2, \theta^{(l,i)}_3)).
    \end{split}
\end{equation}
We expect that this circuit \textit{ansatz} results in $\chi_q\sim 3^{L_1L_2}$, which is exponential with the depth of the VQML model. In Appendix.~\ref{app:bonddim}, we numerically confirm that $\chi_q$ for this \textit{ansatz} averaged over multiple random initializations of $\rvec{\theta}$, indeed exhibits an exponential scaling with respect to the circuit depth. Our numerical observation implies that the typical $\Cq$ of VQML models using polynomially growing computational resources can possess exponentially large bond dimensions, which makes function classes of these VQML models hard to generate classically.


To summarize, every VQML model is an MPS model with a feature map $\mathbf{T} : \rvec{x} \mapsto \mathbf{T}(\rvec{x})$, but subject to a special kind of coefficient MPS determined by the circuit \textit{ansatz}. In the language of ML, VQML models are  MPS model that possesses special \textit{regularization} on coefficient MPS. For classical ML models, regularization is a process to reduce the complexity of learned functions and is executed by adding additional terms in the loss function or by using some heuristics during the training~\cite{Mohri18}. Analogously, employing VQML models corresponds to using an implicit `quantum' regularization by using quantum circuits as coefficient generators. Such a quantum regularization can be implemented classically by contracting a 2D unitary network (quantum circuit) as a coefficient tensor, which is in general hard for classical computers but efficient for quantum. This is where VQML models differ from classical MPS models.

\section{\label{Sec:IV}Dequantization of function classes of VQML models}

The study of quantum-algorithm dequantization aims for developing \textit{efficient} classical algorithm that are of comparable performances to the respective quantum algorithms, both of which utilize the same level of pre-computational power. Following the discussions in Sec.~\ref{Sec:II}, here we define the notion of the VQML-model function.
\begin{definition}[Dequantization of VQML's function class\label{def:dequantization}]
    For a given function class of the VQML model $F_Q(\rvec{x};\rvec{\Theta},\{\phi_\alpha\}_\alpha)$ utilizing $n$-qubits, a set of parameters $\rvec{\Theta}$, and pre-processing functions $\{\phi_\alpha\}_\alpha$ obtained prior to VQML, the function class of the VQML model is \textit{dequantized} by a classical computational model $F_C$ if there exists such a $F_C$ that requires $O(\mathrm{poly}(n,1/\delta,1/\epsilon))$ computational resources and $f_C\in F_C$ such that
    \begin{equation}
        \underset{\rvec{\theta}\sim\mu(\rvec{\Theta})}{\mathbb{P}}\left[\mathcal{D}(f_Q(\rvec{x};\rvec{\theta}),f_C(\rvec{x}))\leq \epsilon\right] \geq 1-\delta.
    \end{equation}
\end{definition}

Here, $\mu(\rvec{\Theta})$ represents the uniform measure over the trainable parameters set. This definition depends on the choice of distance function $\mathcal{D}$. Henceforth, terms like `dequantization of VQML' or `dequantizable VQML models' refer to the dequantization of their function classes. Additionally, throughout this paper, we only consider efficient VQML models that utilize $N = O(\mathrm{poly}(n))$ encoding gates. Therefore $n$ and $N$ can be used interchangeably without altering a computational complexities.

Several points are worth highlighting here. First, it is important to understand that dequantized VQML models may still offer quantum advantages. This is because even if two models belong to the same function class, their performances such as sample efficiency or generalizability can differ largely due to their differences in training landscape. Second, the definition provided above should be considered as a `weak' form. There are scenarios where a trained quantum model generates a non $\epsilon$-approximable function, which resides in a $\delta$ fraction of the function class. Hence, even if a VQML model is $\delta$-dequantized, it may not be possible to classically approximate its other practically relevant output functions.

Nevertheless, the criteria for function class dequantization is a necessary condition for classical surrogates~\cite {Schreiber2023:classical} of VQML models or shadow models of VQML models~\cite{jerbi2023shadows}, the recently investigated. These surrogates or shadow models are classical counterparts designed to efficiently learn from quantum ML models, which can be used to replace the original, \textit{trained} quantum models in later applications. For efficient learning and usage, it is crucial that classical models approximate the quantum model accurately while maintaining efficiency. This is the essential objective of VQML-model dequantization: it ensures that classical models can efficiently approximate their quantum counterparts. 

This dequantization criterion can be extended to general variational quantum algorithms (VQA) beyond just QML scenarios by treating $\rvec{x}$ as variational parameters alongside $\bm{\theta}$. We can determine how `quantum' the functions generated by a given variational quantum circuit family are. This perspective is instrumental in determining the potential of replacing VQA with classical algorithms as a whole. In this context, dequantization of the quantum-circuit \textit{function class} (Def.~\ref{def:dequantization}) is a necessary condition for the overall dequantizability of the VQA itself. In contrast, if a quantum model is \textit{not} dequantizable, then this means it can produce classically inefficient functions and therefore can be a prominent candidate for exhibiting quantum advantage.

We remark that Def.~\ref{def:dequantization} is different with efficient simulatability of quantum circuits. Consider, for instance, the univariate naive Pauli encoded model. This model uses scalar inputs and all $N$ pre-processing functions are identity functions, $\phi_\alpha(x) = x$. Regardless of the circuit's complexity, one can dequantize this model's function class with degree-$N$ Fourier series~\cite{Schreiber2023:classical, Schuld2021:Fourier}, due to its small dimension of function space. Meanwhile, using exponential encoding~\cite{Shin2023}, results in an exponential ($3^N/2-1$)-degree Fourier series, challenging dequantization with a simple strategy that just exploits a finite Fourier series. However, we can cope with this problem in Sec.~\ref{Sec:V}, by exploiting classical MPS models.

\section{\label{Sec:V}Dequantization using classical MPS models}

We learned that the feature map is always expressible in a tensor-product form, and can thus be efficiently generated, given the set of pre-processing functions $\{\phi_\alpha\}_\alpha$. Using this and the fact that VQML models are MPS models, we take a classical MPS~(CMPS)model that is basis-equivalent to the VQML model we would like to dequantize,
\begin{equation}
    f_C(\rvec{x};\rvec{\theta}) \equiv \Cc(\rvec{\theta}) \cdot \mathbf{T}(\rvec{x}),
\end{equation}
where $\Cc(\rvec{\theta})$ is its coefficient part and $\mathbf{T}(\rvec{x})$ is the feature map of the VQML model. The maximum bond dimension of $\Cc(\rvec{\theta})$, denoted as $\chi_c$, may be set arbitrarily. If $\chi_c$ scales exponentially in $N$---which is the number of tensors in $\mathbf{T}$---, then $f_C$ can surely approximate all VQML functions but it is not efficient as contraction complexity for the CMPS model scales $O(\chi_c^2N)$. Therefore we only focus on $\chi_c \sim O(\mathrm{poly}(N))$. 

For distance measure between functions $\mathcal{D}(f_Q,f_C)$, we shall adopt the two-norm squared distance,
\begin{equation}\label{eq:distance}
    \mathcal{D}_2(f_Q,f_C) := \dfrac{1}{\abs{\Omega}}\int_{\Omega}{\abs{f_Q(\rvec{x}) - f_C(\rvec{x})}^2}d\rvec{x}.
\end{equation} 
This is a natural choice if one considers mean squared error, which is a finite approximate version of the two-norm distance, as a performance measure for function regression tasks. Next, we can bound $\mathcal{D}_2(f_Q,f_C)$ using the two-norm distance of $\Cq$ and $\Cc$ from above as follows, 
\begin{equation}\label{eq:distancebound}
\begin{split}
    \mathcal{D}_2(f_Q,f_C) &= \dfrac{1}{\abs{\Omega}}\int_{\Omega}{\left\{(\mathbf{C^q} - \mathbf{C^c})\cdot \mathbf{T}(\rvec{x}) \right\}^2} d\rvec{x}
    \\&= \bra{\Delta}{\dfrac{1}{\abs{\Omega}}\int_{\Omega}{\ketbra{\mathbf{T}(\rvec{x})}d\rvec{x}}}\ket{\Delta}
    \\&\leq \norm{\ket{\Delta}}^2_2\norm{G}_F,
\end{split}
\end{equation} 
where $\ket{\Delta} := \Cq - \Cc$, $G$ is the Gram matrix of $\mathbf{T}_i(\rvec{x})$s, $G_{ij} :=\dfrac{1}{\abs{\Omega}}\int_{\Omega}{\mathbf{T}_i(\rvec{x})\mathbf{T}_j(\rvec{x})}d\rvec{x} $, and $\Omega$ is the domain encompasses all possible inputs, not just the training data. We omitted the trainable parameter, $\theta$, dependence for $\Cc$ and $\Cq$, and one should be aware that they are parametrized independently. From the above inequality, we see that 
\begin{equation}\label{eq:coeffbound}
    \norm{\ket{\Delta}}^2_2 \leq \epsilon/\norm{G}_F,
\end{equation}
which guarantees the approximation within the error tolerance. In other words, good approximability for coefficient tensors in terms of two-norm can be translated to good approximability for functions. Especially, when $\mathbf{T}(\rvec{x})$ contains an orthonormal basis set in a given $\Omega$, such as the Fourier function basis and $\Omega = [-\pi,\pi]$, we have the equality,
\begin{equation}\label{eq:orthonormalequality}
    \dfrac{1}{\abs{\Omega}}\int_{\Omega}{(f_Q(\rvec{x}) - f_{C}(\rvec{x}))^2}d\rvec{x} = \norm{\mathbf{C^q} - \mathbf{C^c}}^2_2 = \norm{\ket{\Delta}}^2_2.
\end{equation}

\subsection{\label{sec:VA}Conditions for not dequantizable VQML models with CMPS models}

\begin{figure*}
    \centering
    \includegraphics[width=\textwidth]{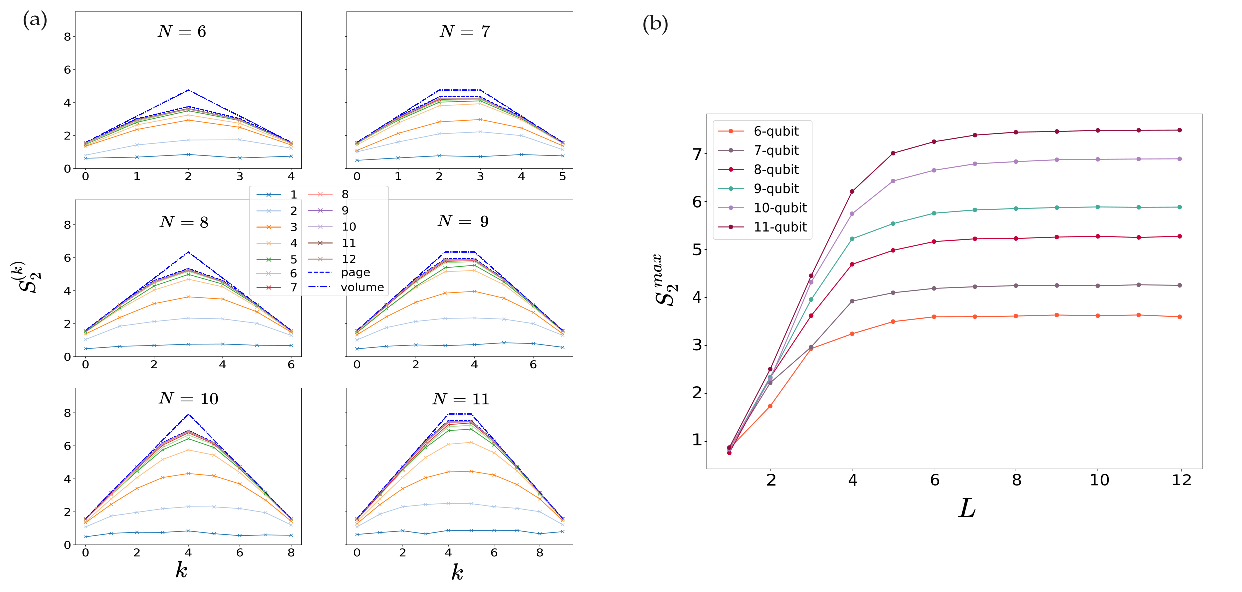}
    \caption{Simulation results for the noiseless simple parallel model case. (a) The Renyi-2 entropy scaling of the coefficient tensors $\mathbf{C^q}$ of the randomly parametrized models is plotted against the subsystem size ($k$). Each average is taken over 30 different parameter sets (24 for $N \geq 10$). The Page curve (blue dotted line) represents the Haar-averaged values for $N$-qutrit quantum states and the volume curve signifies the potential maximum value (the entropy curves for quantum states satisfying the volume law)}. The $S_2^{(k)}$ curves approach the Page curve when the number of layers is sufficiently large ($L \approx N$). (b) The maximum Renyi-2 entropy across all subsystem sizes, $S_2^{max}$, increases with $L$ in the simple parallel model, and saturates at $L \approx N$. 
    \label{fig:simple parallel}
\end{figure*}

From Eq.~\eqref{eq:coeffbound}, and~\eqref{eq:orthonormalequality}, to permit dequantization, an accurate coefficient MPS approximation becomes important. When approximating an MPS $\Cq$ having a maximum bond dimension $\chi_q$ with a restricted MPS $\Cc(D)$ that has a maximum bond dimension $\chi_c = D$, the approximation error in terms of the two-norm is bounded from above by~\cite{Verstraete2006:MPSseminal}
\begin{equation}\label{eq:errorbound}
    \norm{\Cq - \Cc(D)}^2_2 \leq 2\sum_{k=1}^{N-1}\sum_{i=D+1}^{\chi_q}(s^k_i)^2 \equiv 2\eta(D).
\end{equation}
Here, $\{(s^k_i)^2\}_{i=0}^{\chi_q}$ is the set of singular values obtained from the singular value decomposition of $\rho_Q^k:= \Tr_{[k+1,k+2,\dots,N]}{\ketbra{\mathbf{C^q}}}$, which is a reduced matrix for sites $1,2,\dots,k$, and $\eta(D)$ is a truncation error of $\Cq$ which is a sum of discarded singular values when only $D$ largest values are kept.

The magnitude of $\eta(D)$ is dictated by the Renyi-$\alpha$ entropy of $\Cq$'s singular values. Note, also, that $\Cq$ is not a genuine `state', as $\tr{\ketbra{\mathbf{C^q}}} \neq 1$. By properly normalizing the $\ketbra{\mathbf{C^q}}$, one can obtain the Renyi-$\alpha$ entropy for the $k$'th `cut' of $\Cq$,
\begin{equation}
    S_\alpha^{(k)} :=\frac{1}{1-\alpha}\log_2{\tr{(\rho^k_Q)^\alpha}}.
\end{equation}
We focus on the $\alpha = 2$ case as this serves as a criterion for efficient approximability of $\Cq$.

First, we propose a condition for VQML models with all orthonormal basis that are non-dequantizable by CMPS models by virtue of $S_2^{(k)}$,
\begin{proposition}[Highly-entangled coefficient generating models are non-dequantizable by CMPS models.\label{prop:highly-entangled}]
    VQML models satisfying $ G = I$, such that
    \begin{equation}
        \underset{\theta\sim\mu(\rvec{\Theta})}{\mathbb{P}}\left(S_2^{(k)}(\rvec{\theta}) \in O(ck^\beta) \right) \geq 1-\delta
    \end{equation}
    for some constants $c$, and $0<\beta<1$, cannot be dequantized by CMPS models with $\mathcal{D}_2$ distance and any precision $\epsilon$.
\end{proposition}
\begin{proof}
 We use the result from~\cite{Schuch2008:approximabilityofMPS},
\begin{equation} \label{eq:schuch}
    \log{D} \geq S_2^{(k)} +2\log{(1-\epsilon_{tr})},
\end{equation}
where $D$ is the maximum bond dimension of approximating classical MPS $\Cc$, and $\epsilon_{tr}$ is the trace distance between density operators $\ketbra{\Cq}$ and $\ketbra{\Cc}$. 
$G = I$ implies that all basis functions in $\{\mathbf{T}(\rvec{x})_i\}$ are orthonormal, from Eq.~\eqref{eq:orthonormalequality}, we have
\begin{equation}
    \mathcal{D}_2(f_Q,f_C) = \norm{\Delta}^2_2 \leq2\eta(D) \leq 2(N-1)\epsilon_{tr},
\end{equation}
where the second inequality comes from the fact that $\eta(D)$ is bounded by the trace distance. Setting $\epsilon_{tr} = \frac{\epsilon}{2(N-1)}$, we see that
\begin{equation}
    D \geq 2^{S_2^{(k)} + 2\log{\left(1-\frac{\epsilon}{2(N-1)}\right)}},
\end{equation}
for $\mathcal{D}_2(f_Q,f_C)\leq \epsilon$. This states that if $S_2^{(k)}$ of $\Cq$ exhibits a growth rate faster than logarithmic in relation to its size ($N$ or $k = c N$ for $0<c\leq1$), approximating it with efficient CMPS (having $D = \chi_c = O(\mathrm{poly}(N))$ model is impossible for any specified error tolerance $\epsilon$.
\end{proof}

Proposition~\ref{prop:highly-entangled} highlights that the `quantum' nature---classical infeasibility for an approximate generation---of a VQML output function is due to a highly entangled coefficient in an exponentially large dimensional space.

Besides the scaling of $S_2^{(k)}$, we may also look at $S_2^{max}:= \underset{k}{\max}~S_2^{(k)}$. This stems from the reasoning that when an $\chi_c \gg 2^{S^{max}_2} $, the CMPS model is expected to provide a suitable approximation~\cite{noh2020}. Consequently, we regard a VQML model having a larger $S^{max}_2$ as a \textit{harder} model to dequantize compared to one with a smaller $S_2^{max}$. We provide numerical evidence indicating that $S_2^{max}$ can be a good estimator for efficient approximability of $\Cq$ in Appendix~\ref{app:trunerror}. 

From Sec.~\ref{Sec:IIIC} and Appendix~\ref{app:bonddim}, $k$-th bond dimension $\chi_q^{(k)}$ of $\Cq$ can grow exponentially with respect to $N$. As $S_2^{(k)} \leq \log_2{\chi^{(k)}_q}$, $S_2^{(k)}$ can also scale linearly with the site index $k$, and $S_2^{max}$. Therefore, we expect typical VQML models that are sufficiently deep to be hard to dequantize. In the following, we numerically confirm when highly entangled $\Cq$ is generated, thereby finding numerical evidence of non-dequantizable VQML models. 



\subsubsection{Noiseless VQML models\label{sec:noiseless}}

Firstly we investigate the scaling behavior of $S_2$ of typical $\Cq$s from simple parallel models (Fig.~\ref{fig:vqml models}(b.1)) by varying the number of qubits $N$ and number of layers $L$ per trainable unitary ($W_1$ to $W_2$). Here, we set $L = L_1 = L_2$, because when $L_1+L_2$ is fixed, setting $L_1=L_2$ gives us the largest entanglement. (See Appendix~\ref{app:C}). 
 
The result is given in Fig.~\ref{fig:simple parallel}. It is observed that the $S^{(k)}_2$ curve approaches the Page curve for the Renyi-2 entropy \cite{Fujita2018:pagelimit} as $L$ increases, where saturation occurs when $L \approx N$. The Page curve exhibits an almost linear scaling with respect to the subsystem size $k$, and its maximum value (at $k = \lfloor N/2 \rfloor$) exhibits a faster-than-logarithmic scaling with respect to $N$. Specifically, $S_2^{max}(L) \sim 0.79N$ upon saturation. Consequently, these simulation results indicate that for polynomial-depth VQML models, typical $\Cq$ can possess high entanglement efficiently with a small number of parameters, so they are not likely to be dequantizable. Meanwhile, we can observe that shallow circuit depths generally lead to a sufficiently small $S_2^{max}$, so that opens a possibility for dequantization by CMPS models.
 
Secondly, we study the data re-uploading model~\cite{Perez2020}, where the data encoding part is distributed across the trainable parts [Fig.~\ref{fig:vqml models}(b.2)]. As we argued in Sec.~\ref{sec:IIIA}, general encoded VQML models can be rephrased as re-uploading models so that their analysis results are a representative example for general encoded models. We follow Appendix~\ref{app:A} to construct the $\Cq$s for these cases. Especially we compare the basis-equivalent simple parallel and data re-uploading models.

\begin{figure}
    \centering
    \includegraphics[width = 0.5\textwidth]{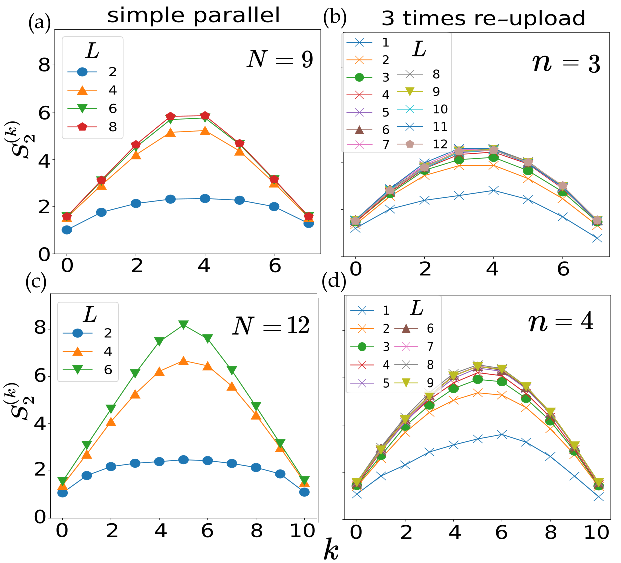}
    \caption{Comparison between basis-equivalent simple parallel and data re-uploading (general encoded) models. We tested $n = 3, 4$ data re-uploading models having $4$ trainable and $3$ encoding blocks. These models (Fig.~(b) and (d)) utilize $3n = N = 9, 12$ Pauli-Z encoding gates, rendering them basis-equivalent to the $n = N = 9, 12$ simple parallel models (Fig.~(a) and (c)) respectively.} The same color lines indicating the same total number of trainable layers, and identical markers denoting the same number of free parameters.\label{fig:compare}
\end{figure}

Figure~\ref{fig:compare} shows the simulation results. The re-uploading model saturates to a lower $S_2^{max}$ compared to the corresponding basis-equivalent parallel model. This indicates that when using the same number of data-encoding gates with enough trainable layers, it is harder to dequantize parallel models. One may also compare two basis-equivalent models that share similar quantum resources such as the total number of trainable layers or the total number of free parameters. Simulation results show that when the parallel model is shallow (L=2 for our case), the re-uploading model is harder to dequantize, but as $L$ increases, the parallel model is always harder to dequantize for the same amount of quantum resources. Therefore, if one desires a VQML model that is not dequantizable, using a parallel model is the more plausible option. 

This conclusion is also a consequence of structural differences between parallel and re-uploading models. When both are basis-equivalent models using $N$ encoding gates, the data re-uploading model uses an $n$-qubit circuit that is smaller than $N$. Therefore, despite employing universal trainable \textit{ansatze}---which can generate any $U \in \mathcal{U}(2^{n})$---for \textit{all} trainable unitary blocks in the data re-uploading model, it is clear that the effective $2^N$-dimensional unitary lies only in a subset of $\mathcal{U}(2^N)$ when the re-uploading model is transformed into simple parallel form using wire-bending techniques (See Fig.~\ref{fig:AppA}). On the other hand, the simple parallel model can generate any $2^N$-dimensional unitary in $\mathcal{U}(2^N)$ with any universal \textit{ansatz}. This result is consistent with the study in~\cite{casas2023multidimensional}.
 
\subsubsection{Noisy VQML models}

\begin{figure*} 
    \centering
    \includegraphics[width=1\textwidth]{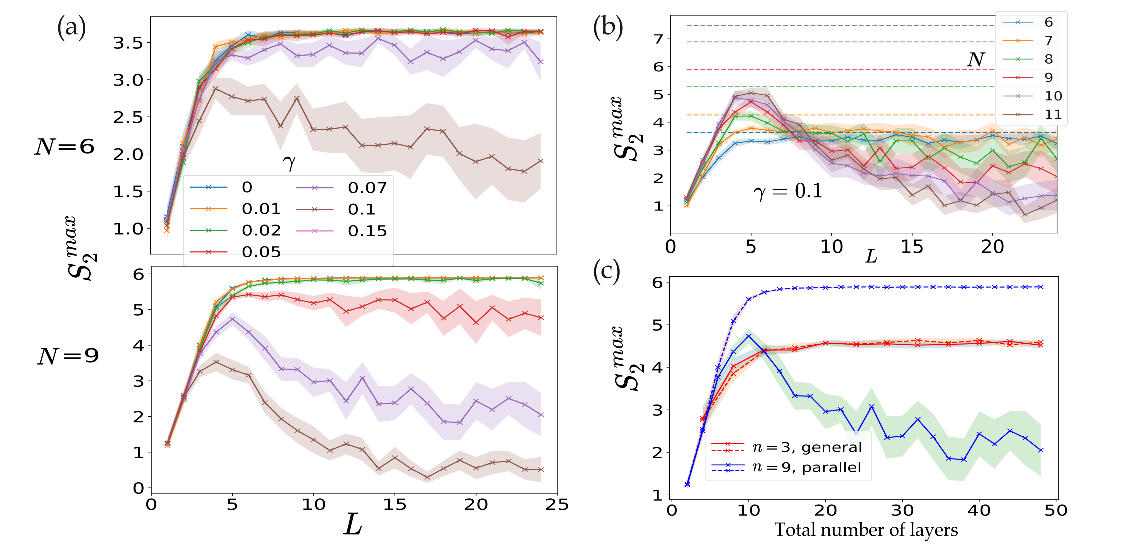}
    \caption{(a) The impact of noise on $S_2^{max}$ for N= 6 and 9 simple parallel models with varying error rates ($\gamma$).  (b) The effect of noise on $S_2^{max}$ for the $\gamma = 0.1$ case with different numbers of qubits. The dashed lines denote the maximum values for noiseless scenarios. (c) A comparison of the $n=N=9$ parallel model and the basis-equivalent $n=3$ re-uploading models when $\gamma = 0.1$. The dashed lines correspond to noiseless scenarios. As the two models possess different numbers of trainable blocks, the x-axis is set to the total number of layers rather than $L$, which indicates the number of layers for a single trainable block. All lines in this figure represent averages over 30 distinct parameter initializations, and the shaded regions indicate the 0.95 confidence level.\label{fig:noisy}}
\end{figure*}

For NISQ devices, every bit of noise counts. Finite noise levels, however small, would destroy coherence, and weaken the entangling power of the circuit, making them easy to simulate with classical computers~\cite{noh2020,Oh2021:lossyboson, Oh2023:tensor, Liu2023:complexity}. Similarly, noise effectively reduces the entanglement in the $\Cq$ of NISQ VQML models, possibly allowing for a dequantization with CMPS models. 

We shall now study the effects of noise by considering noisy two-qubit gates, each of which results in the two-qubit depolarizing error channel
\begin{equation} 
    E(\rho) = (1-\gamma)\rho + \dfrac{\gamma}{4}I\otimes I,
\end{equation}
where $\gamma$ is the error rate. The noisy quantum circuit layer introduces $(1-\gamma)$ factors to the Pauli coefficients of $O'$ and $\rho$. As a consequence, after a sufficient number of noisy layers, $O'$ and $\rho^{\top}$ ultimately become proportional to the identity matrix, so that $\Cq$ converges to a product MPS. We conduct numerical simulations demonstrate the rate at which the entanglement of $\Cq$ decreases while noise is present. Details concerning the simulation and analysis of noisy VQML models can be found in Appendix~\ref{app:noisy}

In Fig.~\ref{fig:noisy}(a), we observe that noise can significantly wash away the entanglement of $\Cq$. For sufficiently large $\gamma$, the influence of noise overwhelms the entangling power of the circuit, resulting in a decrease in $S_2^{max}$ after a relatively small $L$. By comparing the $N=6$ and $N=9$ cases in Fig.~\ref{fig:noisy}(a) and comparing models of different sizes with fixed $\gamma$ in Fig.~\ref{fig:noisy}(b), we observe that larger $N$ experiences more severe noise effects, with significantly large entropy differences. For instance, when $N = 6$, the decrease in $S_2^{max}$ is not particularly significant until $L=24$, whereas when $N=11$, $S_2^{max}$ quickly tends to 1 quickly as $L$ approaches $24$. This finding suggests that, in situations where considerable noise is anticipated, employing a re-uploading model would generate a larger variety of $\Cq$. Fig.~\ref{fig:noisy}(c) supports the aforementioned claim through a comparison between the basis-equivalent ($n=3$), three-round re-uploading model and $N=9$ parallel model. In the absence of noise, the parallel model is harder to dequantize. Still, when noise is present (with $\gamma=0.1$ in this instance), the re-uploading model becomes harder to dequantize as the total number of layers in trainable circuits increases.

\subsection{Dequantizable VQML models.}

Thus far, we proposed a sufficient criterion for non-dequantizable VQML models by CMPS models, namely that when their coefficient MPS $\mathbf{C_q}$ possess sufficiently high entanglement. Here we claim a more general and stronger statement regarding dequantizable VQML models.
\begin{proposition}\label{prop:dequantizable}
    VQML models that have $O(\mathrm{poly}(N))$ linearly independent functions in the basis set can be dequantized by CMPS models.
\end{proposition}

\begin{proof}
    The basis functions of VQML models are components of the feature vector $\mathbf{T}(\rvec{x})$. Suppose the set $\{\mathbf{T}_i(\rvec{x})\}_i$ contains $K$ linearly independent functions, indexed by $i' \in I$. Then, every function in the  VQML model's function class can be represented as:
    \begin{equation}
        \sum_{i' \in I} C_{i'}\mathbf{T}_{i'}(\rvec{x}),
    \end{equation}
    for some vector $C = \sum_{i' \in I} C_{i'}\ket{i'}$. Each computational basis ket $\ket{i'}$ is a product vector as it contains only one non-zero element. Summing MPSs results in a linear increase in bond dimension at most~\cite{Schollwock2011:dmrg}. Consequently, $C$ has a bond dimension of at most $K$. This demonstrates that all functions in a VQML model with $K$ linearly independent basis functions can be constructed using a CMPS model with $\Cc$ of bond dimension at most $K$. The computational complexity of this model is $O(K^2N)$, which proves the proposition.  
\end{proof}

Proposition~\ref{prop:dequantizable} is `strong' in the sense that it applies to \textit{all} functions (all $\rvec{\theta}$ in $\rvec{\Theta}$) in the function class. Moreover, this dequantizability only comes from the property of the feature map, not depending on how complex or hard $\Cq$s from the VQML models are. In other words, VQML models having polynomial-dimensional function class can be dequantized with $\delta = 0$, irrespective of the specific choice of the distance function $\mathcal{D}$ in definition~\ref{def:dequantization}, and regardless of the circuit \textit{ansatz} employed. Proposition~\ref{prop:dequantizable} shows that it is necessary to have an exponential dimension for a genuine non-dequantizable model. We have presented the naive Pauli-encoded model as a representative example of a dequantizable model, and now we can confirm this fact by treating it as a special case of the above proposition.

To demonstrate proposition~\ref{prop:dequantizable} more extensively, we conduct function regression tasks using CMPS models with $f_Q(\rvec{x}) =\Cq \cdot \mathbf{T}(\rvec{x})$s as the target function. The target coefficient $\Cq$ is generated by a noiseless $N=6,8$, $L=10$ simple parallel model. We normalize $\Cq$ so that $\norm{\Cq}_2 = 1$. Such models result in a high $S_2^{max} = 3.61, 5.29$, suggesting the need for a CMPS having $\chi_c \gg 2^{S_2^{max}}$ for small $\norm{\Delta}_2^2$. We choose two different encoding strategies, naive encoding [$\phi_{\alpha}(x) = x$], and exponential encoding [$\phi_{\alpha}(x) = 3^{\alpha-1}x$].

\begin{figure*}
    \centering
    \includegraphics[width = 0.95\textwidth]{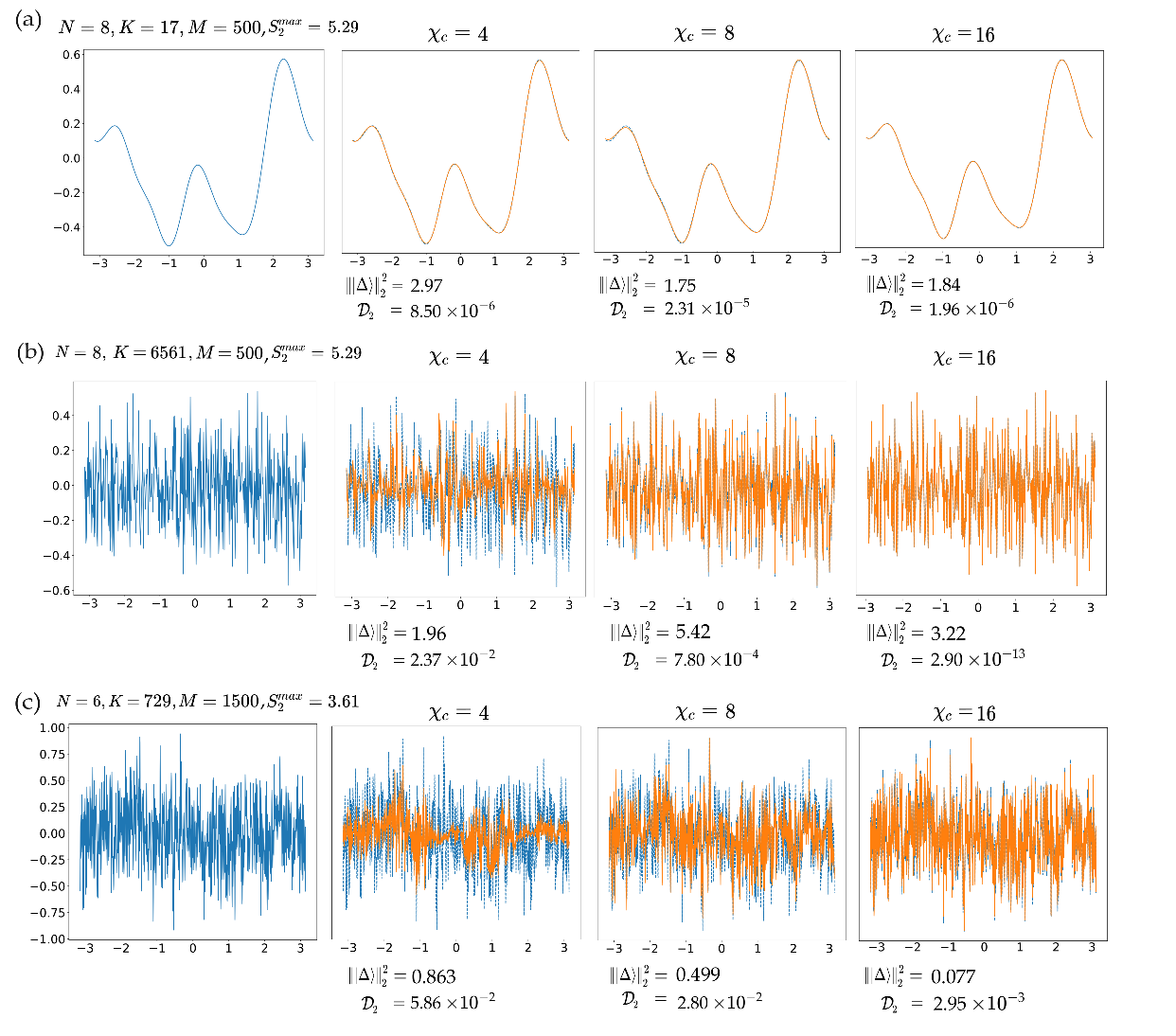}
    \caption{The VQML function regression with CMPS models. The target functions (Blue lines), $f_Q(x)$, are generated by randomly parameterized $L=10$ VQML models in noiseless settings. The variable $K$ represents the number of linearly independent basis functions of the models. The training set is composed of $\{x_j, f_Q(x_j)\}_{j=1}^M$, where $x_j$ values are linearly spaced numbers ranging from $-\pi$ to $\pi$. The optimization of $\mathbf{C^c}$s was carried out with all $M$ data points serving as training data. The orange lines depict the values from the denoted $\chi_c$-CMPS models after 500 training epochs. Each graph provides the 2-norm distance of coefficients after training, $\norm{\ket{\Delta}}^2_2$, and function distance $\mathcal{D}_2$ between the target and approximating functions. The naive Pauli encoding was used in (a) to generate $poly(N)$ dimensional function, while in (b) and (c), exponential encoding is employed. Target functions of (a) and (b) share the same coefficient tensor $\mathbf{C^q}$, which have the same $S^{max}_2$ values.}
    \label{fig:function_fit}
\end{figure*}

The number of linearly independent functions in the components of $\mathbf{T}(\rvec{x})$---the dimension of the function space---is equal to the rank of the Gram matrix $G$. Therefore, one can say that VQML models can be dequantized by $\text{rank}(G)$ CMPS models (if rank${(G)}$ is polynomial to $N$). It is important to note that the rank of $G$ is also affected by the total input domain $\Omega$ which encompasses all possible inputs including training and test datasets. If $\Omega$ is a discrete set having $M$ elements, then $\text{rank}(G) \leq M$, which says the effective dimension of the function class is limited to $M$. This is because every component in $\mathbf{T}(\rvec{x})$ can now be represented by an $M$-dimensional vector. In other words, if we have limited access to only $\mathrm{poly}(N)$ input points (note that we are \textit{not only} considering training points but also all possible test inputs), then VQML models are dequantizable by CMPS models.

Since $\text{rank}(G) \propto \norm{G}_F$, one can also infer from Eq.~\eqref{eq:distancebound} that a lower rank of $G$ allows a large variation in the coefficients, suggesting that VQML models with a smaller rank of $G$ are more susceptible to dequantization, as they allow for larger coefficient discrepancies.

Fig.~\ref{fig:function_fit}(a) depicts the approximation performances using small bond-dimensional CMPS models. The target functions are from $N=8$ naively-encoded VQML models, resulting in rank$(G)$ being polynomial in $N$ ($K = 2N+1$). In this scenario, it is expected that a $\chi_c = K = 17$ would perfectly express the VQML target functions. However, in this case, even a CMPS model with $\chi_c = 4$ would have been sufficient to closely approximate the target function.

Fig.~\ref{fig:function_fit}(b) represents the scenario where we employ exponential encoding to generate functions of exponentially large dimensions. However, $M = 500 \ll K = 3^8$, so rank$(G) \leq 500$ is significantly lower than the maximum possible value.

 We note high similarity between $\Cc$ and $\Cq$ is not required for a close approximation of the function in these cases. One can also observe the unproportionate relationship between $\norm{\ket{\Delta}}^2_2$ and $\mathcal{D}_2$ in certain situations. These simulation results tell us that CMPS models with very large bond dimensions are not necessary for VQML dequantization when rank($G$) is small---CMPS models can approximate functions from VQML models well despite the hard-to-approximate $\Cq$s (high $S_2^{max}$) when rank$(G) \ll O(exp(N))$. As a side note, in Fig.~\ref{fig:function_fit}(c) we use sufficiently many sample points to achieve full-rank $G$. In this case, one witnesses the proportionate relationship between function and coefficient distances, which requires small $\norm{\Delta}_2^2$ for small $\mathcal{D}_2$ as expected form the inequality in Eq.~\eqref{eq:coeffbound}.

In summary, in this section, we identified conditions for which VQML function classes are (likely) dequantizable or not by CMPS models:
\begin{itemize}
    \item VQML models with rank$(G) = O(\mathrm{poly}(N))$ (where the dimension of the function space is $O(\mathrm{poly}(N))$ are dequantizable. 
    \item Noiseless shallow-depth (logarithmic in $N$) circuit models are likely dequantizable, while poly-depth circuits are not.
    \item Noisy VQML models possessing large widths and depths are likely dequantizable.
\end{itemize}
Note that while the first statement is rigorous, the second and third are based on numerical evidence corresponding to Figs.~\ref{fig:simple parallel},\ref{fig:compare} and \ref{fig:noisy}. So informally, a non-dequantizable VQML function requires a function space of an exponentially large dimension and a highly entangled coefficient on this function space.

\section{\label{Sec:VI}Efficient classical kernel induced from a quantum kernel.}

The kernel method plays a crucial role in the context of FLM.~\cite{hofmann2008kernel} Every feature map $\mathcal{F} : \mathbb{R}^d \rightarrow \mathbb{R}^K$ in FLM introduces a kernel $\mathcal{K}(\rvec{x}_i,\rvec{x}_j) = \bra{\mathcal{F}(\rvec{x}_i)}\ket{\mathcal{F}(\rvec{x}_j)},$ which is the inner-product evaluation between feature-mapped data. Quantum kernel methods utilize quantum circuits to generate elements of kernel matrices. Analogously, a quantum kernel is defined as
\begin{equation}
    \mathcal{K}_q(\rvec{x}_i,\rvec{x}_j) = \Tr{\sigma(\rvec{x}_i)\sigma(\rvec{x}_j)},
\end{equation} 
where $\sigma(\rvec{x}) = \mathbf{S}_{enc}(\rvec{x})\ketbra{0}\mathbf{S}^{\dag}_{enc}(\rvec{x})$ is the data-encoded quantum state~\cite{Schuld2021quantumkernel}. For any given quantum kernel, we have pre-processing functions that were pre-computed before implementing $\mathbf{S}_{enc}$, and this naturally induces (normalized) a basis-equivalent product kernel 
\begin{equation}\label{eq:productkernel}
\begin{split}
    \mathcal{K}_{c}(\rvec{x}_i,\rvec{x}_j) &= \dfrac{1}{2^N}\braket{\mathbf{T}(\rvec{x}_i)}{\mathbf{T}(\rvec{x}_j)} \\
    &= \dfrac{1}{2^N}\prod_{\alpha=1}^N[1+\text{cos}(\phi_\alpha(\rvec{x}_i)-\phi_\alpha(\rvec{x}_j))],
    \end{split}
\end{equation} 
where $\mathbf{T}$ is constructed from the pre-processing functions of the given quantum model. Note that it takes $O(N)$ time steps to calculate Eq.~\eqref{eq:productkernel} as it simply involves an inner product between two tensor-product MPSs.

The representer theorem~\cite{Mohri18} states that any optimal function $f^{\text{opt}}$ that minimizes a given empirical regularized loss functional $\mathcal{L} : (f, \{x_i, y_i\}_{i}^{M_t}, \lambda) \rightarrow \mathbb{R}$ can be represented as 
\begin{equation}
    f^{\mathrm{opt}}(\cdot) = \sum_{i=1}^{M_t} \gamma_i\mathcal{K}(\cdot,x_i).
\end{equation}
The optimal weights $\rvec{\gamma}=(\gamma_1,\gamma_2,\ldots,\gamma_{M_t})^\top$ admit analytical solution when we know the whole kernel matrix elements evaluated with the training dataset $\{x_i, y_i\}_{i}^{M_t}$. The function $f^{\mathrm{opt}}$ resides in the so-called reproducing kernel Hilbert space (RKHS)~\cite{Mohri18}, which is the function space that is spanned by the kernel functions $\mathcal{K}(\cdot, x_i)$s. 

\begin{figure}[t]
     \includegraphics[width = 0.95\columnwidth]{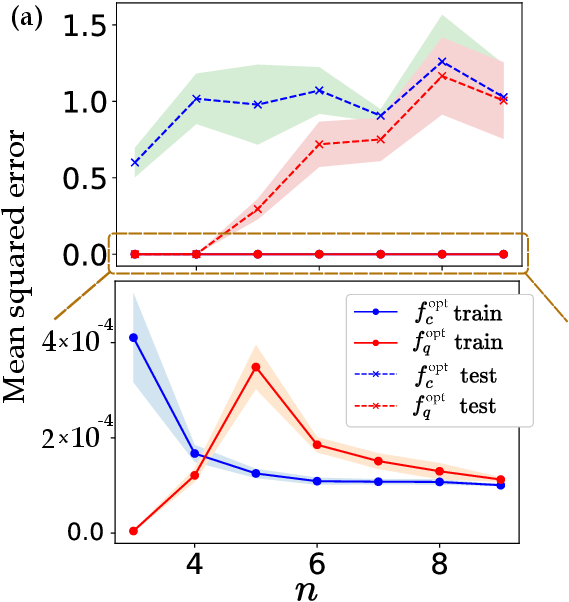}
     \caption{\label{fig:kernel} Training and test losses derived from the kernel methods, which are trained using the VQML generated relabeled f-MNIST dataset. The figure panel offers a magnified view of the training losses.}
\end{figure}

From the observation that every function generated from a quantum circuit encoded with classical data can be represented as an FLM with feature map $\mathbf{T}$, we have the following proposition.

\begin{proposition}\label{prop:kernel}
    The RKHS from any quantum kernel $\mathcal{K}_q(\rvec{x}_i,\rvec{x}_j) = \Tr{\sigma(\rvec{x}_i)\sigma(\rvec{x}_j)}$ using $N$ pre-processing functions $\{\phi_\alpha\}_\alpha$ is included in the RKHS of efficient basis-equivalent product kernel
    \begin{equation}
        \mathcal{K}_c(\rvec{x}_i,\rvec{x}_j) =  \bra{\mathbf{T}(\rvec{x}_i}\ket{\mathbf{T}(\rvec{x}_j)}.
    \end{equation}
\end{proposition}
\begin{proof}
    Let a function from the RKHS of $\mathcal{K}_q$ be 
    \begin{equation}
    \begin{split}
        f_q(\rvec{x}) &= \sum_i a_i\Tr{\sigma(\rvec{x}_i)\sigma(\rvec{x})}\\
        &=\Tr{O\sigma({\rvec{x}})},
    \end{split}
    \end{equation}
    where $O = \sum_i a_i \sigma(\rvec{x}_i)$. 
    Noting that $\sigma(\rvec{x})$ is generated by utilizing data-encoding gates and some non-parametrized quantum gates, this is simply a linear sum of basis functions in $\{\mathbf{T}(\rvec{x})_i\}_{i=1}^{3^N}$, just like a VQML model using the same data-encoding gates. Considering the functions in the RKHS of $\mathcal{K}_c$ are given by
    \begin{equation}
        f_c(\rvec{x}) = \mathbf{C}\cdot\mathbf{T}(\rvec{x}),
    \end{equation}
    we conclude the proof.
\end{proof}

It is important to note that proposition~\ref{prop:kernel} is not about the dequantization of RKHS of quantum kernels in the sense of definition~\ref{def:dequantization}, but rather about the inclusion between two RKHSs of different kernels. Now let us denote the RKHS from quantum/classical kernel as $\text{RKHS}_{q/c}$. The optimal function $f^{\mathrm{opt}}$ from the quantum kernel method indeed belongs to $\text{RKHS}_q \subseteq \text{RKHS}_c$. However, the existence of the analytical solution $\rvec{\gamma}$ of $f^{\mathrm{opt}}$ posits the absence of any constraint on the coefficients of the model. That is when $f^{\mathrm{opt}}$ is represented in the MPS form, $f^{\mathrm{opt}}(x) = \rvec{c}^{\mathrm{opt}}(\rvec{\gamma}) \cdot \mathbf{T}(\rvec{x})$, the optimal coefficient vector $\rvec{c}^{\mathrm{opt}}(\rvec{\gamma})$ can be \textit{any} $3^N$-dimensional vector that has a large bond dimension when represented as an MPS. In other words, some functions in $\text{RKHS}_q$ might not allow for an efficient classical description using a poly-$\chi_c$ CMPS model, and is thus non-dequantizable in general (unless it is $O(\mathrm{poly}(N))$-dimensional such that proposition~\ref{prop:dequantizable} holds). 

\begin{figure*}[ht]
    \centering
    \includegraphics[width = 1.0\textwidth]{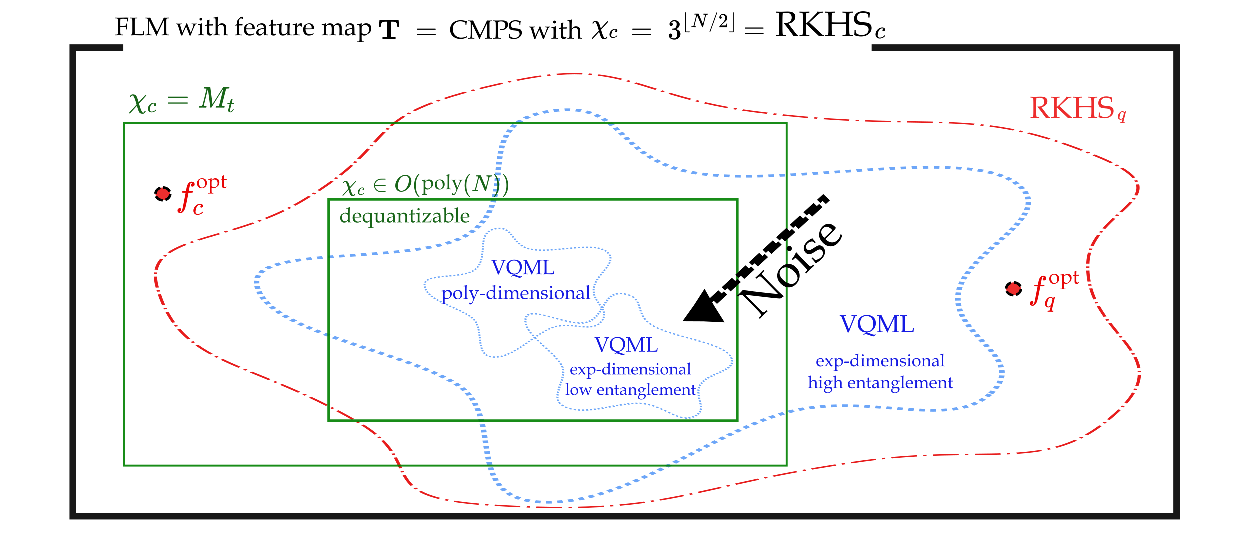}
    \caption{The relationship between the function spaces of basis-equivalent CMPS models, VQML models, and the corresponding RKHSs. All models reside within the MPS model with the feature map $\mathbf{T}$, where the maximum bond dimension is $3^{\lfloor N/2\rfloor}$. The blue wavy stars denote the function spaces of VQMLs, and the green rectangles indicate the space of CMPS models with designated $\chi_c$. Noise can reduce the entanglement of $\Cq$, thereby enabling dequantization and reducing the size of the function space. The $f^{\mathrm{opt}}_{c/q}$ are optimal functions derived from the corresponding kernel methods. Furthermore, $f^{\mathrm{opt}}_{c}$ trained with $M_t$ data is strictly contained within the CMPS model space of $\chi_c = M_t$.}
    \label{fig:Ben}
\end{figure*}

As discussed above, RKHS includes parametrized function classes, so that constrained $\mathbf{C^q}$s also belong to RKHS$_q$. Moreover, upon noticing that,
\begin{equation}
    f^{\mathrm{opt}}_{c} = \sum_{i=1}^{M_t} \gamma_i\mathcal{K}_c(\rvec{x},\rvec{x}_i) = \dfrac{1}{2^N}\sum_i^{M_t} \gamma_i\bra{\mathbf{T}(\rvec{x}_i)}\ket{\mathbf{T}(\rvec{x})},
\end{equation}
according to the representer theorem, we find that $\dfrac{1}{2^N}\sum_i^{M_t} \gamma_i\bra{\mathbf{T}(\rvec{x}_i)} = \Cc_{\mathrm{opt}}$ is an MPS with bond dimension at most $M_t$, which is the number of training data. Equivalently, $f^{\mathrm{opt}}_{c}$ resides in the function class of CMPS models having $\chi_c = M_t$. The relationships among the function classes of VQML, CMPS, and RKHSs are illustrated in Fig. \ref{fig:Ben}. The reader may consult Ref.~\cite{Jerbi2023:beyondkernel} for more discussion about the difference between the variational model and the kernel method.

Kernel methods sharing the same RKHS do not necessarily exhibit the same performances such as generalizability. Here, we compare the quantum kernel method based on IQP encoding using two repetitions of encoding gates, which is conjectured to be hard to simulate classically~\cite{Havlicek2019}, and the corresponding basis-equivalent product kernel method. We again consider the function regression task from the relabeled f-MNIST dataset as in Sec.~\ref{sec:mpsgen}, but now following Refs.~\cite{Huang2021power, Jerbi2023:beyondkernel}, target values are generated by the randomly parametrized $n$-qubit quantum circuits. The IQP encoding of repetition two contains $2n^2$ $\phi_\alpha$s when transformed to a parallel model \textit{via} the procedure in Appendix~\ref{app:A}. These $2n^2$ pre-processing functions include the trivial function $\phi_\alpha(\rvec{x}) = 0$, and the basis-equivalent CMPS model has a length of $2n^2$. However, as $\phi_\alpha(\rvec{x}) = 0$ does not affect the number of linearly independent basis functions, there are only $4n-2$ non-trivial pre-processing functions. We order them as
\begin{equation}
    \phi_\alpha(\rvec{x})=\begin{cases}&x_\alpha,~~~ \alpha \in [1,n] \\
    &x_{\alpha - n} x_{\alpha - (n-1)},~~~ \alpha \in [n+1,2n-1] \\
    &x_{\alpha-(2n-1)},~~~ \alpha \in [2n,3n-1] \\
    &x_{\alpha - (3n-1)} x_{\alpha - (3n-2)},~ \alpha  \in [3n,4n-2],
    \end{cases}
\end{equation}
and create the $\mathbf{T}(\rvec{x})$ with $N = 4n-2$ to evaluate $\mathcal{K}_{c}$. We fit models with a sample of $500$ training data using kernel ridge regression which exploits regularized loss Eq.~\eqref{eq:regularizedloss} with $\lambda = 0.01$, and test losses are computed with $100$ unseen data. 

Figure~\ref{fig:kernel} presents the mean squared error (MSE) value from training and test datasets after training. It is evident that the training loss incurred using the kernel method using $\mathcal{K}_c$ is comparable to that of $\mathcal{K}_q$; both are of the order $10^{-4}$. This indicates that $f^{\mathrm{opt}}_c$ fits the training data almost flawlessly, reflecting its good expressivity on VQML-generated functions. The training loss from $\mathcal{K}_c$ is lower than any other classical methods explored in the study by Ref.~\cite{Jerbi2023:beyondkernel}. However, the test loss is higher than the quantum kernel for smaller system sizes, though the two become comparably worse as the system size expanded. 

Poor generalizability of the quantum kernel with increasing quantum-circuit size is expected as explained in Appendix H. in Ref.~\cite{Huang2021power}. The key point is that the kernel matrix approaches the identity as the Hilbert-space dimension grows exponentially in the qubit number so that good generalizability demands an exponentially growing data sample size. The same thing happens in basis-equivalent kernel $\mathcal{K}_c$ as it inherits an exponentially large dimensional feature map. We note that adjusting pre-processing functions can help improve generalization in quantum kernel~\cite{canatar2023:bandwidth}, and this is also applicable to basis-equivalent product kernel if the VQML model is a simple parallel model, as two kernels are the same.

\section{\label{Sec:VII}Conclusions and Discussions}

In this work, by using the tensor network (TN) formalism, we reveal that any variational quantum machine learning (VQML) model using classical data as its input is a linear model in the featured space, but has constrained coefficients and a feature map of tensor-product form. This general structure enables us to treat VQML models as a subclass of matrix product state (MPS) machine learning models, and offers a nuanced understanding of their characteristics in order to distinguish classically accessible components from genuinely quantum aspects. In QML applications we introduced a definition for dequantizing VQML model function classes, establishing it as a necessary condition for substituting VQML algorithms with classical ML algorithms as a whole. Employing classical MPS models for dequantization, we identified conditions under which VQML models are (or are not) dequantizable. Our numerical analysis provides evidence illustrating these conditions. In carrying out kernel learning methods, we propose a basis-equivalent product kernel, that is efficient and comparable in expressiveness to quantum kernels, opening new possible pathways for their dequantization.

The premise of a fair comparison between a quantum algorithm of interest and its classical counterpart is the consideration of equivalent levels of pre-computations as free operations. Therefore, comparing all models of a common feature map, or that are equivalent in the feature basis is not only a viable approach but also concretely separates what classical models can/cannot perform resource efficiently. Under this basis-equivalent comparison framework, the dequantizing classical model can take any tensor-network structure that could resemble the common multi-scale entanglement renormalization \emph{ansatz}~(MERA), the projected entangled-pair state~(PEPS), or even a neural-network structure that possesses the VQML model's pre-processing functions as its activation functions. An interesting direction for a follow-up study would be to explore the approximation capabilities of these different classical models.

Based on our findings of this work, we see that the core difference between classical and QML models lies in the entanglement content of their coefficient parts, not in the exponentially large feature space as the latter can always be efficiently computed classically. A genuine `quantum' function, which is inefficient to generate with classical models, is one that has a highly entangled coefficient with a small number of parameters in an exponentially large dimensional function space. We suggest looking for data that are attributed to this kind of functions for a quantum advantage with VQML. Recent studies on TN structured QML ~\cite{Huggins2019towards, Araz2022:TNbasedQMLonLHCdata, Du202:PQCExpressive, Haghshenas2022:VariPower, rieser2023tensor} is in line with this perspective: they utilize quantum circuits instead of the dense classical tensor blocks to construct the TN model, resulting in a highly-entangled yet sparsely-parametrized network. These kinds of procedures may be understood as an implicit `quantum' regularization of variational models, which, incidentally, is also carried out as soon as one chooses VQML models over classical MPS ones. Indeed the efficiency in implementing quantum regularization is the advantage of quantum devices. We conducted performance comparisons in Appendix~\ref{app:performance} between VQML and CMPS models, highlighting the better generalizability of quantum regularization on MPS models.

We again emphasize that the dequantization concept discussed in this work relates to the expressivity of machine-learning function classes and does not directly correlate with other aspects of dequantization, such as generalizability or sample complexity during training (trainability). While this study only focuses on the expressivity of VQML models, we recognize the significance of generalizability or trainability issues as well. We believe that the analysis we have developed provides a foundation for further exploration into these crucial aspects of QML. For example, the generalization error bound of ML models is related to the capacity of function classes~\cite{Vapnik1998, Mohri18}, and several capacity measures have been proposed~\cite{Caro2021:encodingdependent, Caro2022:QMLgeneralization, abbas2021power}. Our work could provide an alternative route to access the capacity of the function space through MPS expressivity or operator spreading~\cite{nahum2018:operatorspreading} which relates to the model coefficients.

Regarding the trainability issue, recent research~\cite{Cerezo2023:Doesprovable} has established a link between the trainability and classical simulatability of variational quantum algorithm (VQA) models. They introduced the concept of classical simulation (CSIM) of VQA models, which bears resemblance to our definition of dequantization~\ref{def:dequantization}. While our dequantization only necessitates the \textit{existence} of approximating classical functions in the dequantizing classical model, CSIM demands the approximation of the VQA model's output, given $\rvec{\theta}$ and its description. According to Proposition~\ref{prop:dequantizable}, any model with a poly-dimensional function class can be dequantized. However, if $\Cq(\rvec{\theta})$ is derived \textit{via} a deep quantum circuit, it might be hard to approximate the model's output, based solely on the quantum circuit's description. This scenario highlights a potential divergence between CSIM and dequantization as discussed in this work. Bridging the concepts presented in Ref.~\cite{Cerezo2023:Doesprovable}, which employs the language of operator space, with those of our study---articulated in the language of function space---could yield deeper insights for resolving trainability issues.

We note that in Ref.~\cite{Ballarin2023:entanglemententropy}, extensive results on the scaling and growth of entanglement entropy in VQML models using the MPS formalism are discussed. While their findings may appear to be similar to our simulations in Sec.~\ref{sec:VA}, our objects of analysis are different. Reference ~\cite{Ballarin2023:entanglemententropy} focuses on the entanglement entropies of \emph{quantum states} in VQML circuits, whereas we investigate the entanglement entropies of \emph{coefficient MPS}s describing the featured linear models. This distinction is crucial, as entanglement at the function level can significantly differ from circuit-level entanglement. Our approach not only relates circuit-level entanglement to function but also identifies non-critical entanglement in states, as discussed in Appendix~\ref{app:coefficients}. Moreover, we note that ML models utilizing TN structures have been extensively studied in the literature but without any direct connection to QML~\cite{Stoudenmire2016supervised, novikov2017exponential, Liu2023TNunsupervised, Efthymiou2019:TML, Liu2019:machine, Reyes2021:Muliti-scale}, the latter of which is now uncovered by the main results of this work. The fruits of labor for such a connection are a unified TN perspective on VQML models and the concept of function-class dequantization, which enables a deeper understanding of hardness in variational quantum models.


\acknowledgments
The authors are grateful for insightful and beneficial discussions with C. Oh, H. Kwon, and C. Y. Park. Also thanks to R. Sweke for giving insightful comments. This work is supported by Hyundai Motor Company, the National Research Foundation of Korea (NRF) grants funded by the Korea government (Grant Nos. 2023R1A2C1006115, RS-2023-00237959, NRF-2022M3E4A1076099, and NRF-2022M3K4A1097117) via the Institute of Applied Physics at Seoul National University, and the Institute of Information \& Communications Technology Planning \& Evaluation (IITP) grant funded by the Korea government (MSIT) (IITP-2021-0-01059 and IITP-2023-2020-0-01606). Y.S.T. acknowledges support from the Brain Korea 21 FOUR Project grant funded by the Korean Ministry of Education.

\appendix

\section{VQML models are MPS models} \label{app:A}

\subsection{Simple parallel case}
\begin{figure*}[t]
    \centering
    \includegraphics[width = \textwidth]{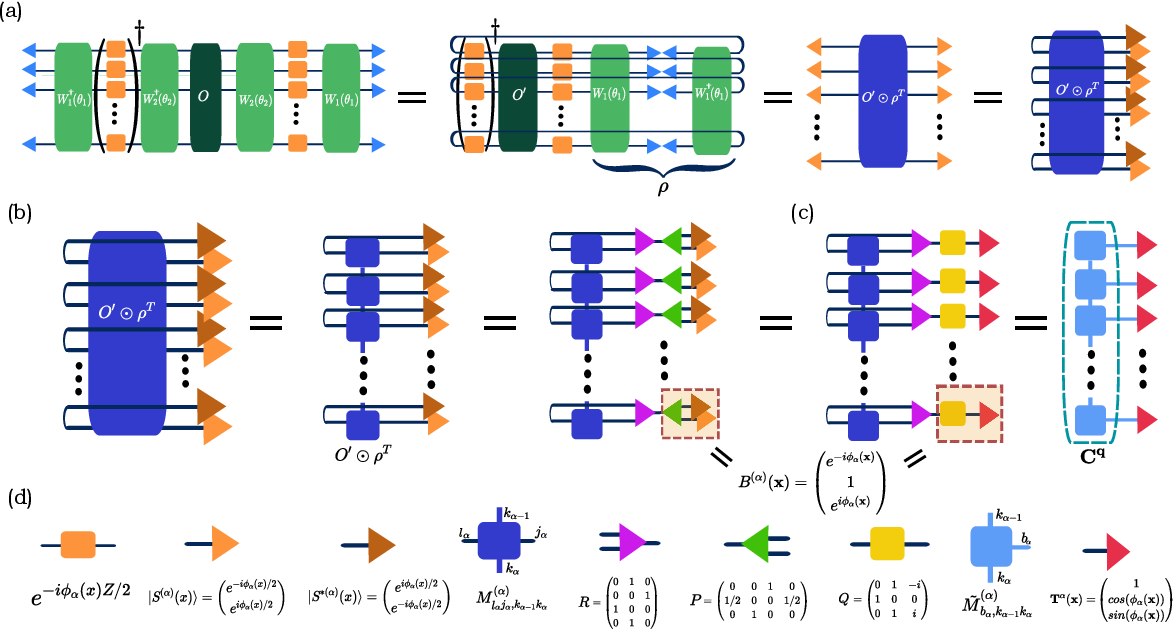}
    \caption{Graphical description of transforming a simple parallel VQML model to MPS model form. (a) The model, $f_Q(\rvec{x}; \rvec{\theta})$, into an FLM form. For the sake of simplicity, we have omitted the $\rvec{\theta}$ dependence in both $O'$ and $\rho^T$. (b) The coefficient part, denoted as $O'\odot\rho^T$, is morphed into an MPS form. This process involves transforming it into a Matrix Product Operator (MPO), vectorizing it, and applying additional tensors. The resulting MPS model form incorporates the feature map $\mathbf{B}$. (c) The final result has the feature map that is $\bigotimes_{\alpha=1}^N\mathbf{T}^{(\alpha)}(\rvec{x})$. The coefficient component becomes the MPS $\mathbf{C^q(\rvec{\theta}_1, \rvec{\theta}_2)}$, representing the contracted form of $(O'\odot\rho^T)\cdot \mathbf{R}\cdot\mathbf{Q}$. At this stage, all tensors are real-valued. All the tensors become real-valued. (d) A detailed description of each block is provided. The index for the site is denoted as $\alpha$.}
    \label{fig:TNbending}
\end{figure*}

The simple parallel models are written as
\begin{equation}
    \begin{split}
    f_Q(\rvec{x};\rvec{\theta},W_1,W_2, O) &= \bra{0}W_1^{\dag}(\rvec{\theta}_1)\mathbf{S}^{\dag}(\rvec{x}) W_2^{\dag}(\rvec{\theta}_2) O \cdots \\& W_2(\rvec{\theta}_2)\mathbf{S}(\rvec{x}) W_1(\rvec{\theta}_1)\ket{0}, \label{eq:app:simple parallel}
\end{split}
\end{equation}
where $\mathbf{S}(\rvec{x}) = \Pi_{\alpha=1}^N e^{-i\phi_\alpha(\rvec{x})Z_\alpha/2}$, and $\rvec{\theta} \equiv (\rvec{\theta}_1, \rvec{\theta}_2)$. 
We start from rewriting Eq.~\eqref{eq:app:simple parallel} as
\begin{equation}
    f_Q(\rvec{x};\rvec{\theta}, W_1, W_2, O) = \Tr{\mathbf{S}^{\dag}(\rvec{x})O'(\rvec{\theta}_2)\mathbf{S}(\rvec{x})\rho(\rvec{\theta}_1)},
    \label{eq:trace form}
\end{equation}
 where we denoted the evolved observable as $O'(\rvec{\theta}_2) := W_2^{\dag}(\rvec{\theta}_2) O W_2(\rvec{\theta}_2)$, and pre-encoded state as $\rho(\rvec{\theta}_1) := W_1(\rvec{\theta}_1)\ketbra{0}W_1^{\dag}(\rvec{\theta}_1) $. Observe that $\mathbf{S}(\rvec{x})$ is diagonal, so we use the property of the Hadamard product (denoted by $\odot$)~\cite{horn2012matrix}
 \begin{equation}
     \Tr{D^{\dag}_1AD_2B} = \bra{D_1}(A\odot B^T)\ket{D_2},
 \end{equation}
 where $D_1 (D_2)$ is a diagonal matrix and $\ket{D_1} (\ket{D_2})$ corresponds to the ket constructed from elements of matrix $D_1 (D_2)$. Using this, we have
 \begin{equation}
     f_Q(\rvec{x};\rvec{\theta}, W_1, W_2, O) = \bra{\mathbf{S}(\rvec{x})} (O' \odot \rho^{T})(\rvec{\theta}_1, \rvec{\theta}_2) \ket{\mathbf{S}(\rvec{x})},
 \end{equation}
 where
\begin{equation}\label{eq:hadamard form}
\ket{\mathbf{S}(\rvec{x})} =\bigotimes_{\alpha=1}^N\ket{S^{(\alpha)}(\rvec{x})} =\bigotimes_{\alpha=1}^N \begin{pmatrix} e^{-i\phi_\alpha(\rvec{x})/2} \\ e^{i\phi_\alpha(\rvec{x})/2}
\end{pmatrix}.  
\end{equation}
We are considering $N$-qubit circuit, so $O'\odot \rho^T$ is a $2^N \times 2^N$ matrix having $2N$ indices. We \textit{vectorize} this $(O' \odot \rho^{T})(\rvec{\theta}_1, \rvec{\theta}_2)$ by gathering the same site indices, thereby obtaining
\begin{equation}
    \begin{split}
         f_Q(\rvec{x};\rvec{\theta}, W_1, W_2, O) = \bra{ (O' \odot \rho^{T})(\rvec{\theta}_1, \rvec{\theta}_2)}(\ket{\mathbf{S}^*(\rvec{x})}\otimes\ket{\mathbf{S}(\rvec{x})}).
    \end{split}
\end{equation}
See Fig. \ref{fig:TNbending}(a) for the graphical description. 

As a result, we see that the VQML model is decomposed into the basis part which only depends on the input data and pre-processing functions $\phi_\alpha$s, and the coefficient part which depends on the rest. For a further analysis of the coefficient part, we represent $\bra{(O'\odot\rho^T)(\rvec{\theta})}$ in the MPS form
\begin{equation}
 \bra{(O'\odot\rho^T)(\rvec{\theta})} = \sum_{k_1k_2\cdots k_{N-1}} M_{l_1j_1k_1}^{(1)}M_{l_2j_2k_1k_2}^{(2)}\cdots M_{l_Nj_Nk_{N-1}}^{(N)},
\end{equation}
where all matrices $M^{(\alpha)}$s are parametrized by $\rvec{\theta}$, and an upper index $\alpha$ denotes the `site' of MPS. The index $\alpha$ ranges from $1$ to the number of encoding gates $N$. One should understand $l_\alpha j_\alpha$ as a physical index of $\alpha$'th site of the MPS. The number of qubits in the circuit $n$ equals $N$ for simple parallel models we consider here, but $N \neq n$ for a general structured model as explained in Appendix. \ref{app:A}.

Note that the feature kets $\ket{S^{*(\alpha)}(\rvec{x})}\otimes\ket{S^{(\alpha)}(\rvec{x})}$ have redundant elements, $1$s. To remove this redundancy, we adopt local tensors,
\begin{equation}
\begin{split}
    \mathbf{P}^{(\alpha)}_{ljb} =&~ (\delta_{b0}\delta_{l1}\delta_{j0} + \frac{1}{2}\delta_{b1}(\delta_{l0}\delta_{j0} + \delta_{l1}\delta_{j1}) +\delta_{b2}\delta_{l0}\delta_{j1}) \\
    =&~ \begin{pmatrix}
        0 & 0 & 1 & 0 \\
        1/2 & 0 & 0 & 1/2\\
        0 & 1 & 0 & 0
    \end{pmatrix}, \quad\quad ( b \times lj~~~\text{matrix form}),
    \end{split}
\end{equation}
and 
\begin{equation}
\begin{split}
    \mathbf{R}^{(\alpha)}_{ljb} =&~ (\delta_{b0}\delta_{l1}\delta_{j0} + \delta_{b1}(\delta_{l0}\delta_{j0} + \delta_{l1}\delta_{j1}) +\delta_{kb}\delta_{l0}\delta_{j1}) \\
    =&~ \begin{pmatrix}
        0 & 1 & 0 \\
        0 & 0 & 1 \\
        1 & 0 & 0 \\
        0 & 1 & 0
    \end{pmatrix}, \quad\quad ( lj \times b~~~\text{matrix form}).
    \end{split}
\end{equation}

Using these tensors we have
\begin{equation}
\begin{split}
    &\sum_{l,j}M^{(\alpha)}_{lj}S^{*(\alpha)}(\rvec{x})_{l}S^{(\alpha)}(\rvec{x})_{j} \\= &\sum_{l,j,l',j',b}M^{(\alpha)}_{lj}\mathbf{R}^{(\alpha)}_{ljb}\mathbf{P}^{(\alpha)}_{bl'j'}S^{*(\alpha)}(\rvec{x})_{l'}S^{(\alpha)}(\rvec{x})_{j'} \\ =&\sum_{l,j,b} M^{(\alpha)}_{lj}\mathbf{R}^{(\alpha)}_{ljb}\mathbf{B}^{(\alpha)}_b,
    \end{split}
\end{equation}
where $\mathbf{B}(\rvec{x})$ is defined in Eq.~\eqref{eq:FLM} and Eq.~\eqref{eq:basis}.
This leads to an FLM representation that is compatible with Eq. \eqref{eq:FLM}, 
\begin{equation}
 \begin{split}
     f_Q(\rvec{x};\rvec{\theta}, W_1, W_2, O) 
     &=(O'\odot\rho^T)(\rvec{\theta})\cdot \mathbf{R} \cdot \mathbf{B}(\rvec{x})\\
     &:= \rvec{c}(\rvec{\theta}) \cdot \mathbf{B}(\rvec{x}).
 \end{split}
 \end{equation}
We shall drop the $\alpha$ index when dealing with the $N$ tensor product of $\alpha$-indexed tensors, and $\cdot$ indicates the tensor contraction. See Fig. \ref{fig:TNbending}(b). 

Lastly, as $f_Q$ is a real-valued function, we can switch to real tensors using the identity
\begin{equation}
\begin{split}
    \mathbf{B}^{(\alpha)} &= \mathbf{Q}^{(\alpha)}\mathbf{T}^{(\alpha)} \\
    &\equiv \begin{pmatrix}
        0 & 1 & -i \\
        1 & 0 & 0  \\
        0 & 1 & i  \\
    \end{pmatrix}
    \begin{pmatrix}
        1\\
        \cos{(\phi_\alpha(\rvec{x}))}  \\
        \sin{(\phi_\alpha(\rvec{x}))}  \\
    \end{pmatrix}.
    \end{split}
\end{equation}
By contracting $\mathbf{Q}$ to MPS part, we get MPS model using a trigonometric basis,
\begin{equation}
\begin{split}
     f_Q(\rvec{x}; \rvec{\theta}, O) &= \sum_{\rvec{k},\rvec{b}}\tilde{M}^{(1)}_{b_1k_1}\tilde{M}^{(2)}_{b_2k_1k_2}\cdots\tilde{M}^{(N)}_{b_Nk_{N-1}}\mathbf{T}^{(1)}_{b_1}\cdots \mathbf{T}^{(N)}_{b_N}\\ 
     &:= \sum_{\rvec{b}}\mathbf{C^q}_{b_1b_2\cdots b_N}\mathbf{T}^{(1)}_{b_1}\cdots \mathbf{T}^{(N)}_{b_N} \\
     &:= \mathbf{C^q}(\rvec{\theta}) \cdot \mathbf{T}(\rvec{x}),\end{split}
\end{equation}
where $\tilde{M}^{(\alpha)}_b = \sum_{i,j,b'} M^{(\alpha)}_{ij}\mathbf{R}^{(\alpha)}_{ijb'} \mathbf{Q}^{(\alpha)}_{b'b} :=M^{(\alpha)}_{ij}\mathbf{\tilde{P}}^{(\alpha)}_{ijb}$, and revived the $\rvec{\theta}$ dependence on $\Cq$. This proves the lemma.~\ref{lem:simpleparallel}. Indices $k_j$ are called `bond indices' and range from $1$ to some maximum numbers, which are called `bond dimensions'.

\subsection{General case}
\begin{figure*}
    \centering
    \includegraphics[width = \textwidth]{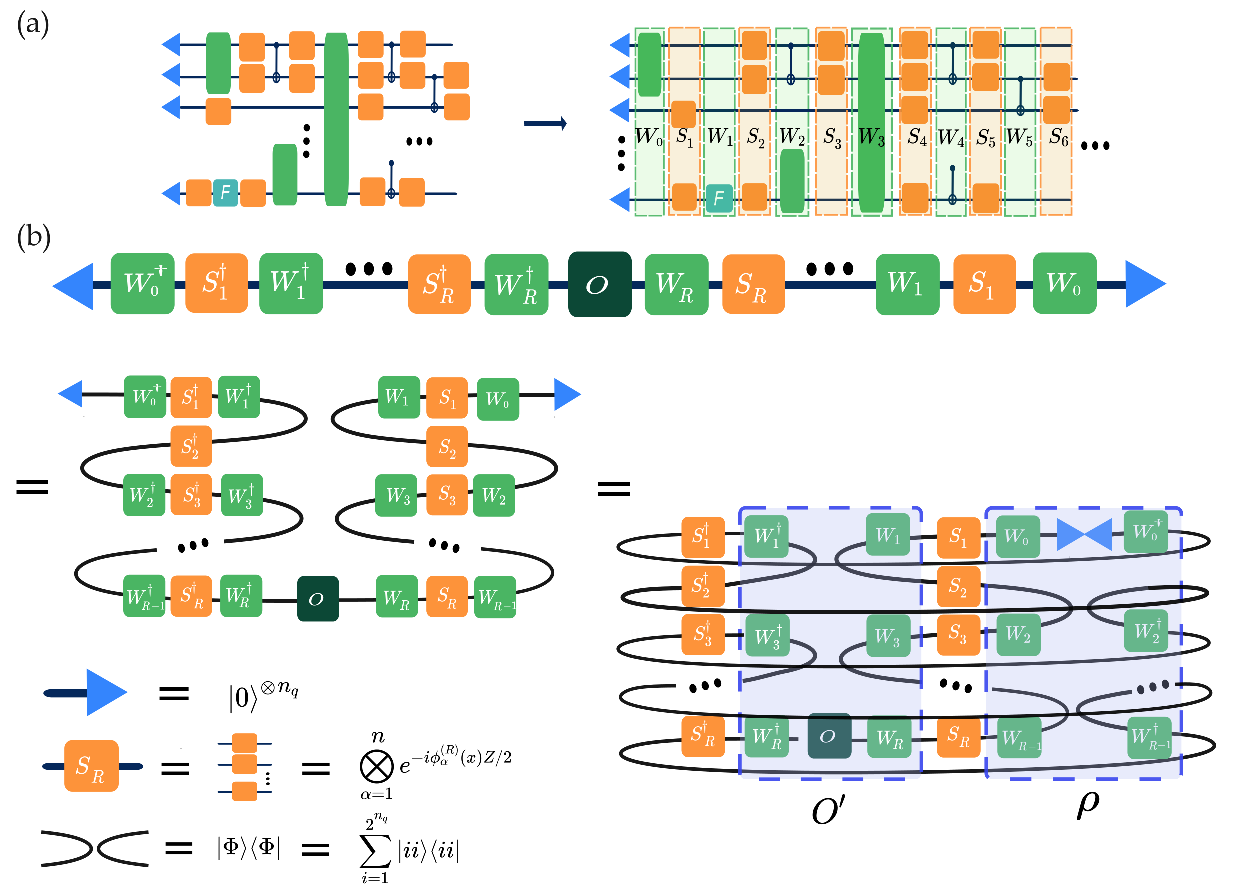}
    \caption{Graphical description of changing the general structure model to MPS form. Orange squares are Pauli-z rotations that are dependent on pre-processed input data, and the others are the coefficient part that does not depend on the input data. For simplicity, we omitted the data dependence and parameter dependence. (a) A Quantum circuit with a general structure is segregated into an alternating encoding part and the coefficient part. As the diagram indicates, the coefficient parts $\mathbf{W}_k$s can contain non-trainable unitaries. (b) After adopting an unnormalized maximally entangled state $\ketbra{\Phi}$, we can transform it as if it were a simple parallel model. By replacing $O'$ and $\rho$ in the main text.}
    \label{fig:AppA}
\end{figure*}
A general encoding strategy can have data-encoding gates throughout the quantum circuit. We decompose all the encoding gates and discriminate the encoding part and trainable part as different layers, resulting in a data re-uploading model that has an alternating structure of encoding parts $\mathcal{S}_k$s and trainable coefficient parts $W_k$s. Let our encoding gates be all decomposed and changed to Pauli-Z rotations so that it becomes $R$ parallel encoding parts as Fig. \ref{fig:AppA}(a). Then quantumly generated function $f_Q(\rvec{x};\rvec{\theta})$ of general model is
\begin{equation}
    \begin{split}
     f_Q(\rvec{x};\rvec{\theta})& = \bra{\rvec{0}}W_0^{\dag}(\rvec{\theta}_0)\mathcal{S}_1^{\dag}(\rvec{x}) W_1^{\dag}(\rvec{\theta}_1)\mathcal{S}_2^{\dag}(\rvec{x})\cdots \\ &S^{\dag}_R(x)W^{\dag}_R(\rvec{\theta}_R) O W_R(\rvec{\theta}_R)\mathcal{S}_R(\rvec{x}) \cdots \\ &\mathcal{S}_2(\rvec{x})W_1(\theta_1)\mathcal{S}_1(\rvec{x}) W_0(\rvec{\theta}_0)\ket{\rvec{0}}.
    \end{split}
\end{equation}
Note that this model is general enough to encompass any VQML model that uses an encoding strategy. By bending wires, we can transform it into the simple parallel VQML model which we have treated in the main text. The graphical description is given in Fig \ref{fig:AppA}(b). For general encoding strategy, encoding block $\mathcal{S}_k$ can contain the identity operator, and this can be simply thought of as $\phi_{\alpha}^{(k)}(\rvec{x}) = 0$. Now the function becomes the same form with Eq. \eqref{eq:hadamard form} in the main text, 
\begin{equation}
    \begin{split}
     f_Q(\rvec{x};\theta)& = \bra{\mathcal{S}(\rvec{x})} (O'_R \odot \rho_R^{T})(\rvec{\theta}) \ket{\mathcal{S}(\rvec{x})},
    \end{split}
\end{equation}
where $O'_R$ and $\rho_R$ is newly defined as
\begin{widetext}
\begin{equation}\label{eq:app:generalO}
    O'_R =\begin{cases}\bigotimes_{k=1}^{(R-1)/2}(W_{2k-1}^{\dag} \otimes I)\ket{\Phi}\bra{\Phi}(W_{2k-1} \otimes I)\otimes W_R^{\dag}OW_R, \quad\quad\text{if R is odd}\\
     \bigotimes_{k=1}^{R/2}(W_{2k-1}^{\dag} \otimes I)\ket{\Phi}\bra{\Phi}(W_{2k-1} \otimes I)\otimes O, \quad\quad\text{if R is even}
 \end{cases}
\end{equation}
\begin{equation}\label{eq:app:generalrho}
    \rho_R =\begin{cases}W_0\ket{0}\bra{0}W_0^{\dag} \otimes \bigotimes_{k=1}^{(R-1)/2}( I \otimes W_{2k}^{\dag} )\ket{\Phi}\bra{\Phi} (I \otimes W_{2k} ), \quad\quad\text{if R is odd}\\
        W_0^{\dag}\ket{0}\bra{0}W_0 \otimes \bigotimes_{k=1}^{R/2}( I \otimes W_{2k}^{\dag} )\ket{\Phi}\bra{\Phi} (I \otimes W_{2k} ),  \quad\quad\text{if R is even}
 \end{cases}
\end{equation}

\end{widetext}
\begin{equation}
    \ket{\mathcal{S}(\rvec{x})} = \bigotimes_{k=1}^R \bigotimes_{\alpha =1}^{n}\begin{pmatrix}
        e^{-i\phi^R_\alpha(\rvec{x})/2}\\ e^{i\phi^R_\alpha(\rvec{x})/2}
    \end{pmatrix}.
\end{equation}
Here $\ket{\Phi} = \sum_{i=1}^{2^{n}} \ket{ii}$, unnormalized maximally entangled state. Note that now $O'$ and $\rho$ are not real observable nor real state. Fig. \ref{fig:AppA}(b) only depicts when $R$ is even, but one can picture an odd case analogously. 

Although the general-structure model can be represented as a parallel model, there is a significant difference in terms of coefficients. In the general-structure model, the operators $O'$ and $\rho$ become $2^{{n}(R+1)}$-dimensional operators ($2^{{n}R}$ for odd $R$ cases). However, they cannot fully exploit the entire space of the given dimensional operator space, as the available free parameters are significantly fewer than what is required for a $2^{{n}(R+1)}$ ($2^{{n}R}$ for odd $R$) dimensional operator space, even if universal unitary \textit{ansatze} are used for all $\{W_k\}_k$. Consequently, the general-structure model possesses a smaller function space compared to the parallel model when they are basis-equivalent.

\section{Comments on coefficients.\label{app:coefficients}}

\begin{figure*}
    \centering
    \includegraphics[width = 0.95\textwidth]{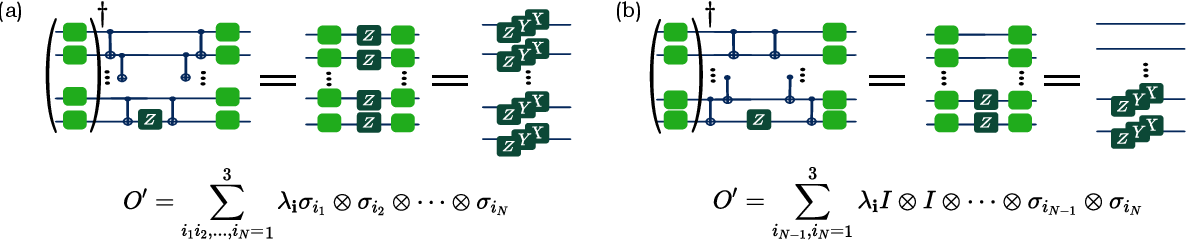}
    \caption{Operator spreading of different \textit{ansatze}. (a) Hadware-efficient \textit{ansatz} used in the main text. This ansatz spread local Z operator on the last qubit to all $N$-qubit space with 1 layer. (b) Hadware-efficient \textit{ansatz} but with the reversed ordering of CNOT gates. This \textit{ansatz} cannot spread local operators to the whole space using only 1 layer. Resulting in only 9 Pauli string coefficients being non-zero.}\label{fig:coeff}
\end{figure*}

In the main text, we adopted tensors 
\begin{equation}
    \mathbf{\tilde{P}} =~ \bigotimes_{\alpha=1}^N\begin{pmatrix}
        1 & 0 & 0 \\
        0 & 1 & i \\
        0 & 1 & -i \\
        1 & 0 & 0
\end{pmatrix},
\end{equation} 
which is a product of $\mathbf{R}$ and $\mathbf{Q}$, to generate MPS from the circuit \textit{ansatz}. This might look a little odd at first sight, but when you write down the $2\times2$ Hermitian matrix $M$ with the Pauli matrix basis as $M^{(\alpha)} = \lambda^{(\alpha)}_0 I + \lambda^{(\alpha)}_1 X + \lambda^{(\alpha)}_2 Y + \lambda^{(\alpha)}_3 Z$, then
\begin{equation}\label{eq:app:coeff}
    M^{(\alpha)}\mathbf{R}^{(\alpha)}\mathbf{Q}^{(\alpha)} = M^{(\alpha)}\mathbf{\tilde{P}}^{(\alpha)}=\tilde{M}^{(\alpha)} =  2\begin{pmatrix}
    \lambda^{(\alpha)}_0 \\
    \lambda^{(\alpha)}_1 \\
    \lambda^{(\alpha)}_2
    \end{pmatrix}.
\end{equation}
Therefore $\mathbf{\tilde{P}}$ discards the Pauli-Z coefficients of $M^{(\alpha)}$s, and multiply 2 to rest of coefficients. Let us represent $O'\odot \rho^T$ with Pauli string basis 
\begin{equation}
    O'\odot\rho^T = \sum_{\rvec{i}} \lambda_{\rvec{i}} ~\sigma^{(1)}_{i_1}\otimes \sigma^{(2)}_{i_2}\otimes \cdots\otimes \sigma^{(N)}_{i_N},
\end{equation} \label{eq:paulicoeff}
where $\rvec{i} \in \{0,1,2,3\}^{\otimes N}$. From the observation above,
\begin{equation}
    \begin{split}
        (O'\odot \rho^T)\cdot\mathbf{\tilde{P}} = 2^N \lambda_{\Tilde{\rvec{i}}},~~~~ \Tilde{\rvec{i}} \in \{0,1,2\}^{\otimes N}. 
    \end{split}
\end{equation}
We see that the coefficient on feature map components $\mathbf{T}_{\Tilde{\rvec{i}}}(\rvec{x})$ corresponds to the $2^N\lambda_{\Tilde{\rvec{i}}}$, which are the Pauli string coefficients of $O'\odot \rho^T$ except the Z containing components.

For instance, let us use exponential encoding on 1d input $x$, where $\phi_\alpha(x) = 3^{\alpha-1}x$ and $N=3$ simple parallel model. One of the basis functions is $\cos(x)\sin(3x)\cos(9x)$. This is chosen by $\Tilde{\rvec{i}} = (1,2,1)$. Therefore the coefficient on basis function $\cos(x)\sin(3x)\cos(9x)$ is the $2^N$ times coefficient on Pauli string $X\otimes Y \otimes X$ when $O'\odot \rho^T$ is represented in Pauli string basis. One might expect high-frequency terms to depend on Pauli strings which have many non-identity elements. However, this is not true in general. Again in the same setting, $\sin(x)\sin(3x)$ depends on $Y\otimes Y \otimes I$ which has 2 non-identity elements. On the other hand, $\cos(9x)$ depends on $I\otimes I \otimes X$ which has 1 non-identity element but has a higher frequency. 

This sheds light on the nature of the coefficients of VQML models, which have been somewhat opaque thus far. It allows us to understand the coefficients \textit{via} knowledge about the operator spreading capability of trainable circuits in the context of Pauli string basis. To see how trainable \textit{ansatz} choice affects the coefficient set, let us look at an example of a simple parallel model, using $W_1(\rvec{\theta}_1) = \bigotimes^N_{\alpha=1} H$. Then $O'\odot \rho^T$ becomes just $\dfrac{1}{2^{N/2}}O'$, where $O' = W^{\dag}_2(\rvec{\theta}_2) O W_2(\rvec{\theta}_2)$ is evolved operator. Next, we set $O = Z_{N-1}$ which is a local Pauli-Z operator, and consider two different cases of $W_2(\rvec{\theta}_2)$s having $L =1$. One is hardware-efficient \textit{ansatz}, which we used throughout the main text. (Fig.~\ref{fig:coeff}(a)) and the other is reversed-CNOT \textit{ansatz} where the ordering of CNOT gates are reversed (Fig.~\ref{fig:coeff}(b)). As CNOT gates spread the Z operator in the target qubit to the control qubit and arbitrary unitary changes Z into an arbitrary superposition of other Pauli matrices (except $I$ that has non-zero trace), $O'$ for the first case possibly possess non-zero $\lambda_{\rvec{i}}$s for all $\rvec{i} \in \{ 1,2,3\}^{\otimes N}$. In Eq.~\eqref{eq:app:coeff}, we saw that Pauli-Z coefficients do not play the role, so it can exploit at most $2^N$ components in $\{\mathbf{T}_{\rvec{i}}(\rvec{x})\}_{\rvec{i}\in\{1,2\}^{\otimes N}}$. However, the second case cannot spread the local Z operator to full qubit space so one gets only $9$ non-zero $\lambda_{\rvec{i}}$s in $O'$. Consequently, the VQML model using the second circuit can only use at most $4$ components out of $3^N$ $\mathbf{T}_{\rvec{i}}$s.

Our $\Cq$ constructions, which utilize Pauli coefficients, capture the easiness of Clifford circuits. Clifford circuits are known to be efficiently simulable by classical computers. Therefore, one might expect that the VQML model which consists of only Clifford gates cannot show any quantum advantage.  Once again, let us set $W_1(\rvec{\theta}_1) = \bigotimes^N_{\alpha=1} H$, but this time, we compose $W_2$ with Clifford gates and set
\begin{equation}
    O = \mathcal{P} \in \bigotimes_{j=1}^N\sigma_j,~~\sigma_j \in{I,X,Y,Z}.
\end{equation}
Clifford circuits merely permute the Pauli coefficients of a given operator, therefore $O'$ simply becomes another Pauli string $ \mathcal{P}'$, which contains only one non-zero element in its coefficient. This implies that our Clifford model uses just \textit{one} basis function. If $O$ had $K$ non-zero Pauli coefficients, then $O'$ could be expressed as the sum of $K$ Pauli strings. This sum would generate an MPO, and consequently, a $\Cq$ with a bond dimension at most $K$. As a result, the function class from this Clifford model can be reproduced by a CMPS model with a bond dimension of at most $K$, which is efficient if $K$ scales polynomially. Although the Clifford circuit, consisting of a finite gate set, doesn't fit into the VQML model, its analysis and the discussion of the previous paragraph hint at how we can link the magic of quantum circuits or operator spreading to the capabilities of VQML models.

\section{\label{app:bonddim}Maximum bond dimension scaling of simple parallel models.}

Two-qubit entangling gates such as CNOT gates can be represented by a matrix product operator (MPO) of bond dimension equal to 2. Contracting two MPOs of bond dimensions $\chi_1$ and $\chi_2$ results in an MPO of bond dimension at most $\chi_1\chi_2$. Therefore, the maximum bond dimension for $O'$ and $\rho^T$ scales at most exponentially with the number of layers $L_1$ and $L_2$. Due to the inequality, rank$(A\odot B)\leq \text{rank}(A)\text{rank}(B)$, maximum bond dimension for $O' \odot \rho^T$ is bounded by min$(4^{\lfloor N/2\rfloor}, 4^{L_1L_2})$. The VQML coefficient vector $\Cq$ is obtained by contracting $\mathbf{\tilde{P}}$ to the $O' \odot \rho^T$, which shrinks the physical dimension from 4 to 3. Therefore, $\chi_q$ can possibly scale as $\sim 3^{L_1L_2}$, which is exponential with the depth of the VQML model.

To see how bond dimension scales with real circuit, we have used the circuit ansatz of Fig.~\ref{fig:vqml models}(b). After the construction of $\Cq$s, we canonicalized them and extracted the number of non-zero elements in every bond indices. The maximum number of non-zero elements is then taken. Results are in Table \ref{tab:chiq}, all averaged over 24 different parameter sets. We set $L_1 = L_2 = L$. The average values of $\chi_q$s scale with $L$ as $4^L$, and quickly saturate to the possible maximum value, $3^{\lfloor N/2 \rfloor}$. This shows that bond dimensions scale maximally in general, thereby coefficient sets of poly-depth VQML models cannot be generated efficiently by the classical MPS, even on average.

\begin{table}
\caption{ Maximum bond dimension for $\mathbf{C^q}$ : $\chi_q$}
\resizebox{\linewidth}{!}{%
\begin{tabular}{>{\centering\hspace{0pt}}m{0.296\linewidth}|>{\centering\hspace{0pt}}m{0.033\linewidth}>{\centering\hspace{0pt}}m{0.05\linewidth}>{\centering\hspace{0pt}}m{0.067\linewidth}>{\centering\hspace{0pt}}m{0.067\linewidth}>{\centering\hspace{0pt}}m{0.067\linewidth}>{\centering\hspace{0pt}}m{0.067\linewidth}>{\centering\hspace{0pt}}m{0.067\linewidth}>{\centering\hspace{0pt}}m{0.067\linewidth}>{\centering\hspace{0pt}}m{0.067\linewidth}>{\centering\arraybackslash\hspace{0pt}}m{0.067\linewidth}} 
\toprule
\diagbox{\textbf{N}}{~\textbf{L}} & \multicolumn{1}{>{\hspace{0pt}}m{0.033\linewidth}}{\textbf{ 1}} & \multicolumn{1}{>{\hspace{0pt}}m{0.05\linewidth}}{\textbf{ 2}} & \multicolumn{1}{>{\hspace{0pt}}m{0.067\linewidth}}{\textbf{ 3}} & \multicolumn{1}{>{\hspace{0pt}}m{0.067\linewidth}}{\textbf{ 4}} & \multicolumn{1}{>{\hspace{0pt}}m{0.067\linewidth}}{\textbf{ 5}} & \multicolumn{1}{>{\hspace{0pt}}m{0.067\linewidth}}{\textbf{ 6}} & \multicolumn{1}{>{\hspace{0pt}}m{0.067\linewidth}}{\textbf{ 7}} & \multicolumn{1}{>{\hspace{0pt}}m{0.067\linewidth}}{\textbf{ 8}} & \multicolumn{1}{>{\hspace{0pt}}m{0.067\linewidth}}{\textbf{ 9}} & \multicolumn{1}{>{\hspace{0pt}}m{0.067\linewidth}}{\textbf{ 10}} \\ 
\hline
\textbf{ 6} & 4 & 27 & 27 & 27 & 27 & 27 & 27 & 27 & 27 & 27 \\
\textbf{ 7} & 4 & 27 & 27 & 27 & 27 & 27 & 27 & 27 & 27 & 27 \\
\textbf{ 8} & 4 & 59 & 81 & 81 & 81 & 81 & 81 & 81 & 81 & 81 \\
\textbf{ 9} & 4 & 59 & 81 & 81 & 81 & 81 & 81 & 81 & 81 & 81 \\
\textbf{ 10} & 4 & 64 & 243 & 243 & 243 & 243 & 243 & 243 & 243 & 243 \\
\textbf{ 11} & 4 & 64 & 243 & 243 & 243 & 243 & 243 & 243 & 243 & 243 \\
\bottomrule
\end{tabular}
}\label{tab:chiq}
\end{table}

\section{Different number of layers for trainable blocks in the parallel model.\label{app:C}} 

\begin{figure}
    \centering
    \includegraphics[width = 0.5\textwidth]{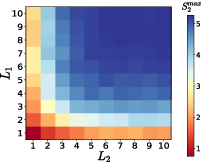}
    \caption{\label{fig:diffL}$S_2^{max}$ for $N=8$ parallel VQML model. $L_1$ ($L_2$) is the number of layers in $W_1(\theta)$ ($W_2(\theta)$).}
\end{figure}

In Fig.~\ref{fig:diffL}, we present a simulation result from the $N=8$ parallel model. When the total number of layers $L_1 + L_2$ is the same, $S_2^{max}$ is the largest around when $L_1 = L_2$. Here $S_2^{max}$ is the largest Renyi-2 entanglement entropy of $\Cq = (O'\odot \rho^T)\cdot \mathbf{\tilde{P}}$. For this reason, we stick to setting all the number of trainable layers to be the same when there is no additional mention. 

This is expected because
\begin{equation}
    \text{rank}(A\odot B) \leq \text{rank}(A)\text{rank}(B),
\end{equation}
so when one of the operators ($O' ~\text{or}~ \rho^T)$ exhibits low rank---low entanglement entropy---Hadamard product of them cannot possess high entanglement entropy. A small number of layers implies low entanglement, thus one can expect that large $S_2^{max}$ can be achievable when both numbers of layers are high enough.

\section{\label{app:trunerror}Truncation errors for $\Cq$s }

\begin{figure*}
    \centering
    \includegraphics[width = \textwidth]{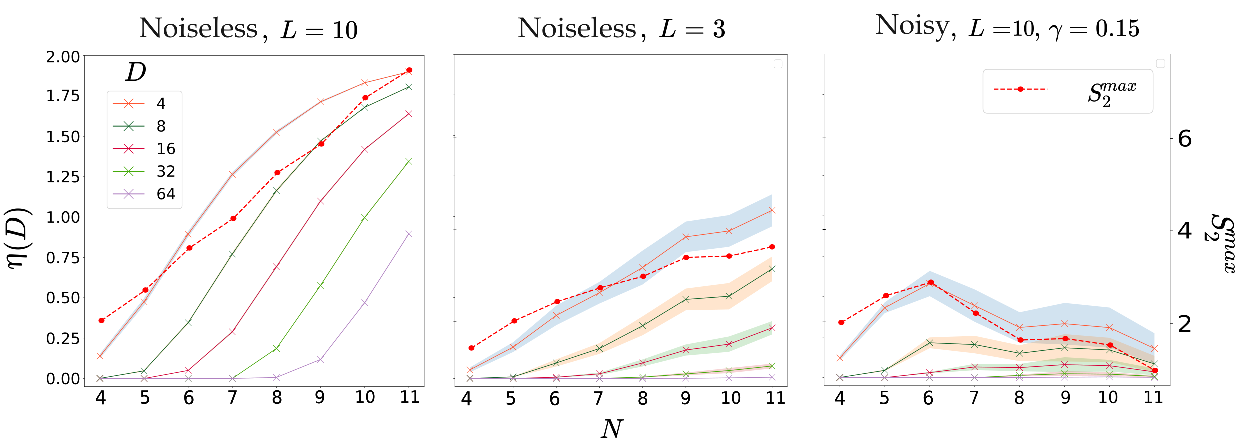}
    \caption{The truncation errors for $\mathbf{C^q}$s when the maximum bond dimension is $D$, alongside the averaged $S_2^{max}$ of original $\Cq$s (represented by red dashed lines), are presented. All values are averaged over 30 different randomly chosen parameter sets.\label{fig:eD}}
\end{figure*}

Let us understand how truncation error $\eta(D)$---which gives us the upper bound of approximation error using $\chi_c = D$ CMPS model---changes with the system size for a given bond dimension bound $\chi_C=D$. Noiseless, $L=10$ parallel VQML models are chosen as ``hard'' models which generate high and linearly-scaling $S^{max}_2$. For ``easy'' VQML models, we chose $\gamma =0.15$, $L=10$ noisy parallel models, and noiseless $L=3$ models. After generating $\Cq$s respectively for each model, we truncated their singular values by leaving only $D$ largest values, with $D=\{4,8,16,32,64\}$ and obtained truncation error $\eta(D)$. 

As in Fig \ref{fig:eD}, the noiseless model requires an exponential increase in $D$ to achieve a fixed truncation error while increasing the model size. Therefore this simulation result indicates that to achieve a fixed approximation error bound using a CMPS while increasing the circuit size, one needs to increase $\chi_c$ exponentially. The noiseless shallow model exhibits a similar trend as the deep model, albeit with considerably smaller absolute values. For noisy cases, the truncation error decreases as the system size increases, and when $N=10,11$ their $\chi_q = 243$ was exponentially large. However, $ D = 2^5 = 32 \ll \chi_q = 243$ was sufficient for almost zero error bound.

These average truncation errors---and hence the approximation error bound---scalings align with the scaling of $S^{max}_2$, as evidenced by the plotted $S_2^{max}$ values. Moreover, note that for nearly zero error bound, $D = 32 \gg 2^{2.56} \sim 5.9$ for the noisy case and $D = 64 \gg 2^{3.63} \sim 12.3$ for the shallow case were enough, where $2.56~(3.63)$ is the maximum $S_2^{max}$ for the noisy (shallow) case. These numerical results demonstrate that $S^{max}_2$ can serve as a reasonable measure for evaluating the ease of approximating the $\Cq$s of VQML models.

\begin{figure*}
    \centering
    \includegraphics[width = 1.0\textwidth]{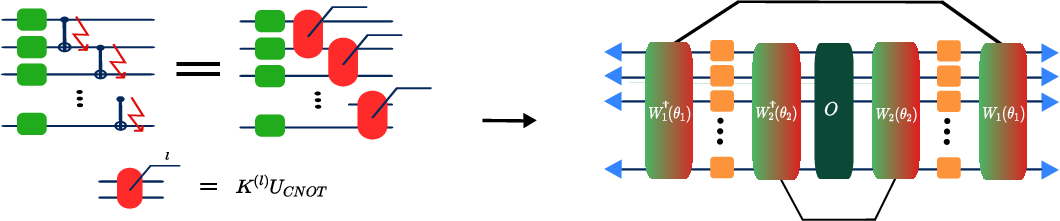}
    \caption{\label{fig:noisycirc}Depolarizing noise acts on after every CNOT gate is applied. $U_{CNOT}$ and Kraus tensor are contracted to create noisy CNOT tensors, denoted as red tensors. The full tensor diagram for the noisy quantum model has additional contraction lines for Kraus sum.}
\end{figure*}

\section{Noisy case analysis.}\label{app:noisy}

\begin{figure*}[ht]
    \centering
    \includegraphics[width = 1.0\textwidth]{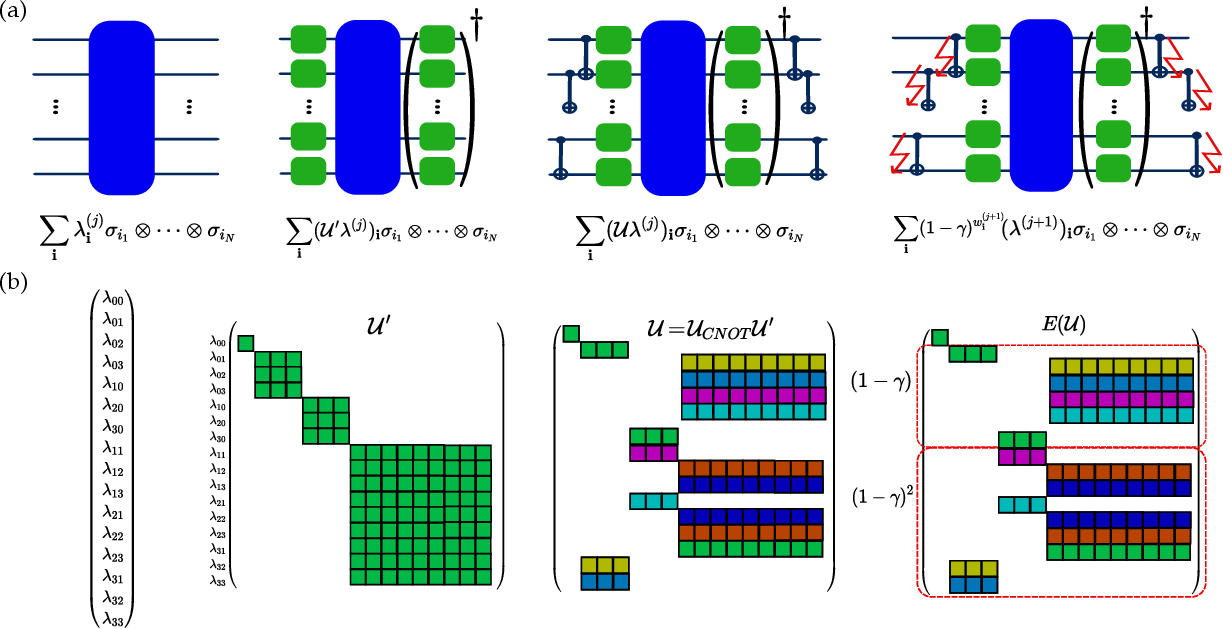}
    \caption{\label{fig:noisy2}(a) Applying one layer of noisy hardware-efficient \textit{ansatz}. (b) This is a graphical depiction of the 2-qubit case Pauli transfer matrix (PTM) for a (noisy) hardware-efficient ansatz. We reordered the Pauli coefficients, depicted as columns of $\lambda_{\rvec{i}}$s, to expose the block-diagonal structure of the PTM for the single-qubit unitaries layer. The PTM of the CNOT gate permutes the order of rows. Permuted rows are denoted in the same color and non-permuted rows are colored in green. Lastly, the multiplied noise factors are indicated. Note that $\lambda_{00}$ is not affected by operations. }
\end{figure*}

For the noisy case, we considered depolarizing error after every two-qubit gate operation. Kraus operators for this error model are given by 
\begin{equation}
    K^{(l)} \in \{ \sqrt{1-\gamma15/16}I\otimes I, \sqrt{\gamma/16} I \otimes X, \cdots, \sqrt{\gamma/16} Z \otimes Z  \}.
\end{equation}
For our simulations, we used CNOT gates for two-qubit operations. Let us denote the state before applying one CNOT gate as $\sigma$. Then after applying CNOT gate and noise channel state becomes
\begin{equation}
    \sigma' = \sum_{l=0}^{15}K^l U_{CNOT}\sigma U^{\dag}_{CNOT}(K^{l})^{\dag}.
\end{equation}
Kraus operator can be understood as a rank-5 tensor where index $l$ is for Kraus sum. We contracted Kraus tensor with $U_{CNOT}$ tensor to construct a rank-5 noisy CNOT tensor. By replacing all the $U_{CNOT}$s in the circuit with a noisy version, and connecting all Kraus indices $1$ to their corresponding conjugated part we could get the noisy version of $\Cq$s. See Fig.~\ref{fig:noisycirc}.

With understanding of coefficients of VQML models as Pauli coefficient, We can analyze how noise affects the coefficients of VQML using the Pauli path integral technique that is introduced in ~\cite{Aharonov2022:polysim, Fontana2023:ClassicSim}. First, we observe that
\begin{equation}
\sum_l K^l(\sigma_i \otimes \sigma_j)(K^l)^\dag = \begin{cases}
    &(1-\gamma)(\sigma_i \otimes \sigma_j),~~\text{if}~ i , j \neq 0\\
    &I\otimes I ,~~\text{if} ~i=j=0.
\end{cases}
\end{equation}

In other words, after the depolarizing channel $E$, all 2-qubit Pauli operators attain $(1-\gamma)$ factor except the Identity. Let us denote observable after applying $j$ (noisy) hardware-efficient \textit{ansatze} as
\begin{equation}
    O^{(j)} = \sum_{i_1i_2,\ldots,i_N} \lambda^{(j)}_{i_1i_2,\ldots,i_N} \sigma_{i_1}\otimes \cdots \otimes \sigma_{i_N}.
\end{equation}
Applying single-qubit unitary $U^{\bigotimes N}(\sigma_{i_1}\otimes \cdots \otimes \sigma_{i_N})(U^\dag)^{\bigotimes N}$ mixes non-identity Pauli matrices while leaving the identity unchanged. Therefore, after applying single-qubit unitaries in $(j+1)$'th layer, we get 
\begin{equation}
\lambda'^{(j)}_{\rvec{i}} = \mathcal{U}'_{\rvec{i}\rvec{i}'}\lambda^{(j)}_{\rvec{i}'},
\end{equation} 
where $\rvec{i} \equiv i_1i_2,\ldots,i_N$, and $\mathcal{U}'$ is the representation of $U^{\bigotimes N}(\cdot)(U^\dag)^{\bigotimes N}$ in Pauli basis or Pauli transfer matrix (PTM) (We have omitted the $\theta$ dependence for simplicity). Note that $\mathcal{U}'$ can be block-diagonalized by simply permuting the order of indices. Next, we apply a CNOT gate. As CNOT gate is an inverse of itself, CNOT gate on $i_k,i_{k+1}$-site qubits exchanges two coefficients in the set $\{\lambda'^{(j)}_{i_1,\ldots,i_k,i_{k+1},\ldots,i_N} | i_{l\neq k, k+1}\text{ are same} \}$. Here index-exchanging follows CNOT change rule which is depicted in Fig.~\ref{fig:noisy2}(b). As a result, we get 
\begin{equation}
\lambda^{(j+1)}_{\rvec{i}} = \mathcal{U}_{CNOT,\rvec{i}\rvec{l}'}\mathcal{U}'_{\rvec{l}'\rvec{i}'}\lambda'^{(j)}_{\rvec{i}'}
\equiv \mathcal{U}_{\rvec{i}\rvec{i}'}\lambda^{(j)}_{\rvec{i}'},
\end{equation}
where $\mathcal{U}$ denotes the PTM of noiseless one layer of hardware-efficient \textit{ansatz}. 

We apply the noisy channel $E$ on $i_ki_{k+1}$-site qubits which introduces $(1-\gamma)$ factor if $i_ki_{k+1} \neq 00$. We do this start from $1,2$-site qubits to $N-1,N$-site qubits resulting in 
\begin{equation}
    (1-\gamma)^{w_{\rvec{i}}}\times \lambda^{(j+1)}_{\rvec{i}}.
\end{equation} Here $w_{\rvec{i}}$ is the number of non-$00$ (non-$II$) sequences in index-vector $\rvec{i}$, or we call it \textit{second-order Hamming weight}, which can range from $0$ to $N-1$. For example, if $\rvec{i} = 002300$, then $w_{\rvec{i}} = 3$. Applying noisy layers from the beginning, the Pauli coefficient $\lambda^{(L)}_{\rvec{i}}$ after $L$-noisy layers is
\begin{equation}
\begin{split}
\sum_{\rvec{i}_0\rvec{i}_1\rvec{i}_2\ldots\rvec{i}_{L-1}} &(1-\gamma)^{w_{\rvec{i}_1}+w_{\rvec{i}_2}+\ldots +w_{\rvec{i}_{L-1}} +w^{(L)}_{\rvec{i}}  }f(\rvec{i}_0\rvec{i}_1,\ldots,\rvec{i}_{L-1};\rvec{i})\\
&\equiv \sum_{\vec{\rvec{i}}_{0:L-1}} (1-\gamma)^{|w|_{\vec{\rvec{i}}_{1:L-1}} + w_{\rvec{i}}}f(\vec{\rvec{i}}_{0:L-1};\rvec{i}),
\end{split}
\end{equation}
where
\begin{equation}
    f(\rvec{i}_0\rvec{i}_1\ldots\rvec{i}_{L-1};\rvec{i}) = \mathcal{U}_{\rvec{i}\rvec{i}_{L-1}}\cdots\mathcal{U}_{\rvec{i}_2\rvec{i}_1}\mathcal{U}_{\rvec{i}_1\rvec{i}_0}\lambda^{(0)}_{\rvec{i}_0}.
\end{equation}
We call the sequence of index-vectors $\vec{\rvec{i}}_{0:L-1}\equiv(\rvec{i}_0,\ldots,\rvec{i}_{L-1})$s as Pauli path as named in Ref.~\cite{Aharonov2022:polysim}, and $|w|_{\vec{\rvec{i}}_{1:L-1}}\equiv w_{\rvec{i}_1} + \cdots + w_{\rvec{i}_{L-1}}$ as total second-order Hamming weight of the Pauli path $\vec{\rvec{i}}_{1:L-1}$.  
Finally, coefficients on basis functions are obtained after the Hadamard product between noisy evolved $O'$ and $\rho^T$. Hadamard product has a mixed product property which is $(A\otimes B) \odot (C \otimes D) = (A\odot C)\otimes(B \odot D)$ and the following product table for Pauli matrices.
\begin{equation}
    \begin{cases}
    I\odot I^T =  Z\odot Z^T &= I,
    \\X\odot X^T = Y\odot Y^T &= X,
    \\-X\odot Y^T = Y\odot X^T &= Y,
    \\I\odot Z^T = Z\odot I^T &= Z,
    \\ \text{otherwise}&=0.
    \end{cases}
\end{equation} Therefore, $\rvec{k}$'th Pauli coefficient of $O'\odot \rho^T$ is
\begin{equation}
\begin{split}
    \lambda^{O'\odot \rho^T}_{\rvec{k}}=\sum_{\rvec{i}\odot \rvec{j} = \rvec{k}}&\sum_{\vec{\rvec{i}}_{0:L-1},\vec{\rvec{j}}_{0:L-1}} (1-\gamma)^{|w|_{\vec{\rvec{i}}_{1:L-1}}+w_{\rvec{i}}}f(\vec{\rvec{i}}_{L-1};\rvec{i})\\
    &\times (1-\gamma)^{|w|_{\vec{\rvec{j}}_{1:L-1}}+w_{\rvec{j}}}f(\vec{\rvec{j}}_{L-1};\rvec{j}).
    \end{split}
\end{equation}
where $\sum_{\rvec{i}\odot \rvec{j}=\rvec{k}}$ denotes that summation over $\rvec{i}$ and $\rvec{j}$ satisfying the condition $\rvec{i}\odot \rvec{j} = {\rvec{k}}$, which has $2^{N}$ combinations.

All Pauli paths except the $(\mathbf{0}, \mathbf{0}, \ldots, \mathbf{0})$ attain noise factors that depend on each paths. As a consequence, $O' \odot \rho^T$ converges to the identity, which becomes product MPS when converted to $\Cq$. We leave a more comprehensive analysis of the noisy case as future research.

\section{Performance of VQML models and CMPS models.\label{app:performance}}

We have seen that highly-entangled coefficiennts make VQML models hard to dequantize. In other words, the unique power of VQML models comes from the ability to \textit{efficiently} generate high-entangled coefficient MPS models using a small number of parameters. In this section, we compare the VQML models and CMPS models in the context of machine learning to explore the performance differences between two models.

All VQML models are simulated and generated classically using Python Pennnylane package~\cite{bergholm2022pennylane}. Tensor contractions and generation of CMPS models are done with the Quimb package~\cite{gray2018quimb}. Optimization of all variational models is done by Adam optimizer with a learning rate 0.01, and 500 training epochs. 

\subsection{Property of coefficients and comparison on function regression}

The number of trainable parameters is a crucial characteristic in machine learning models. Generally, the number of parameters is considered as a measure of the size of the model. More importantly, both models' computational complexities (contraction complexity for CMPS models and gate number complexity for VQML models) are polynomially related to it, thereby setting the same number of parameters enables the comparison between models that share similar computational complexity. Let us set the number of parameters be $P$, then a fundamental difference arises between the two models. In the CMPS model, $S_2^{max} = O(\log{\chi_c}) = O(\log{\sqrt{P/N}})$. On the other hand, for a noiseless VQML model with $L \approx N$, we have $S^{max}_2 = O(N) = O(\sqrt{P})$ as observed in numerical simulations. Consequently, the CMPS model generates a coefficient set characterized by low entanglement and dense parameters, while the VQML model's coefficient set, $\mathbf{C^q}$, exhibits high entanglement and sparse parameters. A dense $\mathbf{C^c}$ can generate any coefficient tensor with maximum bond dimension $\chi_c$, while a sparse $\mathbf{C^q}$ cannot create the majority of coefficient tensors possessing a maximum bond dimension of $\chi_q$. We interpret this as an implicit ``quantum" regularization, which is an efficient process for quantum machines but not for classical MPS models.

To see the performance of implicit `quantum' regularization in MPS models, we conduct a series of regression tasks to compare the performances of basis-equivalent CMPS and VQML models, each equipped with comparable parameter numbers. We set all conditions such as training data, training epoch, optimizer setting, etc, to be the same unless otherwise specified. For the CMPS model, we invoke regularized loss
\begin{equation}\label{eq:regularizedloss}
    \mathcal{L} = \dfrac{1}{M_t}\sum_i (f_c(\rvec{x}_i; \rvec{\theta}) - y_i)^2 + \lambda \norm{\Cc(\rvec{\theta})}^2_2,
\end{equation}
with regularization constants $\lambda \in \{10^{-3},10^{-4},10^{-5},10^{-6},0\}$.

\subsection{Fashion-MNIST with MPS generated label\label{sec:mpsgen}}

\begin{figure}[t]
     \includegraphics[width = 0.95\columnwidth]{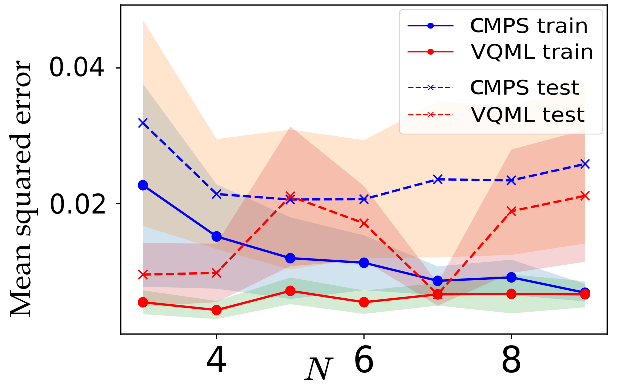}
     \caption{\label{fig:fMNIST} The test and training losses (Mean Squared Error) of the trained models with the MPS-generated labeled f-MNIST dataset are depicted. For the CMPS model, the regularization constant $\lambda = 0$ yields the best performance, and therefore, only results from this case are plotted. The results, which have been averaged over 10 different target coefficient instances, are accompanied by shaded regions representing a 0.95 confidence level.}
\end{figure}

We choose the function regression task from the re-labeled f-MNIST dataset as done in Refs.~\cite{Huang2021power, Jerbi2023:beyondkernel}. Input data are the pre-processed fashion-MNIST data that have dimensions of $n \in [3,9]$. Unlike the previous works, the target values are generated (re-labeled) with the $\chi = 3$ MPS model, so that 
\begin{equation}
\begin{split}
    y_i = \dfrac{1}{K}\sum_{\rvec{b}}^{3}\sum_{\rvec{i}}^{3}M^{(1)}_{b_1,i_1}&M^{(1)}_{b_2,i_1,i_2},\cdots M^{(n)}_{b_n,i_{n-1}}\\
    &\times \mathbf{T}^{(1)}(\rvec{x}_i)_{b_1}\cdots \mathbf{T}^{(n)}(\rvec{x}_i)_{b_n}.
\end{split}
\end{equation}
$K$ is the $\max{\{\abs{y_i}\}_i}$, normalization factor to set the target values lie within the $[-1,1]$. Unlike the quantum circuit-generated case we used simple encoding $\phi_\alpha(\rvec{x}) = x_\alpha$. That is, every element in the vector is encoded once by one Pauli-Z rotation. For the training CMPS model, we employ the same structure CMPS model as the target MPS model, which has $\chi_c = 3$.  For CMPS models, all had $\chi_c= 3$, resulting in $[45, 72, 99, 126, 153, 180, 207]$ free parameters. In parallel, VQML models have $L\in[2,3,3,3,4,4,4,]$ resulting in $[36, 72, 90, 108, 168, 192, 216]$ free parameters. For CMPS model, $S_2^{max} = \log_2{3} \sim 1.7$, whereas the VQML model can have $S_2^{max} > 2$. 

Losses after training are plotted in Fig.~\ref{fig:fMNIST}. Interestingly, while this task is expected to be favored by CMPS models due to their structures being exactly the same as the target-generating MPSs, VQML models show slightly better performance than CMPS models.

\subsection{Step function regression}

\begin{figure*}[t]
    \centering
     \includegraphics[width = 0.95\textwidth]{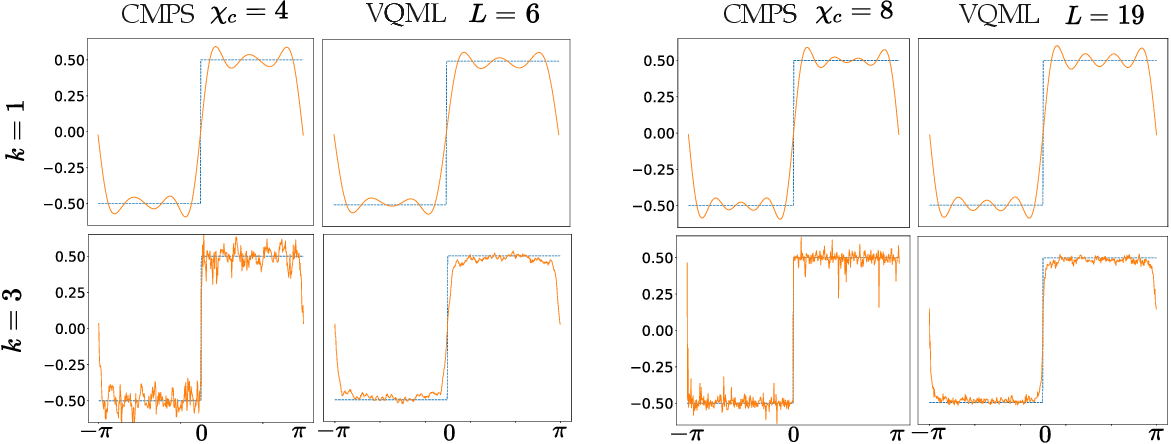}
     \caption{\label{fig:step}A comparison on 1-D step function regression task. The pre-processing functions are given as $\phi_\alpha(x) : k^{\alpha-1}x$. For the CMPS models, we only plot the functions exhibiting the lowest test loss (Mean Squared Error) across all regularizing constants, denoted as $\lambda$. Grouped models illustrate those with a comparable number of free parameters for coefficients. }
\end{figure*}

The step function,
\begin{equation}
    f_{step}(x) = \begin{cases}
        &1/2,\quad \text{if}~ x>0 \\
        &-1/2, \quad \text{if}~ x \leq 0,
    \end{cases}
\end{equation}
is an important function class as it can represent the target function of a classification task. We trained models with $400$ randomly picked data and tested with $100$ unseen data points. For pre-processing functions we chose 
\begin{equation}
    \phi_\alpha(x) = k^{\alpha-1}x,
\end{equation}
where $k \in \{1,3\}$. We compared CMPS models of $N=8$ and parallel VQML models. The CMPS models have $\chi_c = 4~(8)$, resulting in $282~(930)$ free parameters, whreas the VQML models have $ L= 6~(19)$, and $288~(912)$ free parameters. Outputs from trained models are shown in Fig.~\ref{fig:step}. The optimization settings are the same as the variational ridge regression of the re-labeled f-MNIST dataset.

First, when we use naive encoding ($k = 1$), both models show comparable performances. However, when the number of basis functions becomes large ($k = 3$), we can observe slight differences between them. It appears that CMPS models exhibit a stronger tendency to overfit, as indicated by highly spiky graphs. This overfitting behavior becomes more pronounced when $\chi_c$ increases. In our numerical study, the test loss of CMPS model increases from approximately 0.010 to 0.016, while that for VQML models decreases from about 0.013 to 0.009 as we increase the number of free parameters. 

From above regressions, CMPS models seem to suffer overfitting problems more than VQML models, and this might come from the high expressivity of dense MPSs. These show primitive evidence of the advantage of using quantum regularization. We leave comparisons on more different tasks and analytic studies about the generalization ability of quantum regularization as a further research topic. 

\subsection{Comments on computational resources}

We compared the models sharing a similar number of free parameters, so that they have similar computational complexity. However, in practice, real VQML models always accompany statistical error $\delta_q$ and require $O(1/\delta_q^2)$ number of shots, resulting in additional computational resources. Regarding this, we can allow more bond dimensions to CMPS models. To be concrete, an $N$-qubit, poly-depth parallel VQML model uses $O(\mathrm{poly}(N)/\delta_q^2)$ quantum gates while the basis-equivalent CMPS model uses $O(N\chi_c^2)$. Therefore, $\chi_c$ can be $O(\mathrm{poly}(N)/\delta_q)$ for poly-depth quantum models with similar scaling of computational resources. If $\delta_q$ is exponentially small, then an exponentially large bond-dimensional CMPS model is allowed. In this case, VQML loses its advantage in expressivity. 

\subsection{Kernel ridge regression}

For the dataset, We follow the same data pre-processing in \cite{Jerbi2023:beyondkernel}. Original fashion MNIST data is a 28 $\times$ 28-pixel image where pixel values range from 0 to 255. Images are associated with 10 labels. We normalized the pixel values to lie in [0,1] and rescaled them to have 0 mean values. Next, we did Principal Component Analysis (PCA) to reduce the $28\times28$-dimensional vector to $n\in[3,9]$-dimensional vector. 

For the case of kernel method simulation in the main text, we generated the target values $y_i$s using an IQP-encoding first circuit, 
\begin{equation}
    y_i =  \bra{0}S^{\dag}_{IQP}W^{\dag}(\rvec{\theta}_{target})(\rvec{x}_i)Z_1W(\rvec{\theta}_{target})S_{IQP}(\rvec{x}_i)\ket{0},  
\end{equation}
where $\rvec{\theta}_{target}$ is randomly generated and $W(\rvec{\theta}_{target})$ is consists of hardware-efficient \textit{ansatz} with $L$ layers. $k$ is the normalization factor, which is a standard deviation of the training set of $y_i$s. The number of layers $L \in [10,7,6,5,4,4,3]$, so that the free parameters in the target quantum circuit are about 90 parameters. 

Quantum kernels are calculated using Python Pennnylane package~\cite{bergholm2022pennylane}. We use Python package KernelRidge regression in the Scikit-learn package for kernel ridge regression. The regularization constant is set to be $0.01$.

\bibliography{biblo}

\end{document}